# SILKMOTH: An Efficient Method for Finding Related Sets with Maximum Matching Constraints

Dong Deng[*]    Albert Kim[*]    Samuel Madden    Michael Stonebraker

MIT

{dongdeng,alkim,madden,stonebraker}@csail.mit.edu

## ABSTRACT

Determining if two sets are related – that is, if they have similar values or if one set contains the other – is an important problem with many applications in data cleaning, data integration, and information retrieval. For example, set relatedness can be a useful tool to discover whether columns from two different databases are joinable; if enough of the values in the columns match, it may make sense to join them. A common metric is to measure the relatedness of two sets by treating the elements as vertices of a bipartite graph and calculating the score of the maximum matching pairing between elements. Compared to other metrics which require exact matchings between elements, this metric uses a similarity function to compare elements between the two sets, making it robust to small dissimilarities in elements and more useful for real-world, dirty data. Unfortunately, the metric suffers from expensive computational cost, taking $O(n^3)$ time, where $n$ is the number of elements in the sets, for *each* set-to-set comparison. Thus for applications that try to search for all pairings of related sets in a brute-force manner, the runtime becomes unacceptably large.

To address this challenge, we developed SILKMOTH, a system capable of rapidly discovering related set pairs in collections of sets. Internally, SILKMOTH creates a signature for each set, with the property that any other set which is related must match the signature. SILKMOTH then uses these signatures to prune the search space, so only sets that match the signatures are left as candidates. Finally, SILKMOTH applies the maximum matching metric on remaining candidates to verify which of these candidates are truly related sets. An important property of SILKMOTH is that it is guaranteed to output exactly the same related set pairings as the brute-force method, unlike approximate techniques. Thus, a contribution of this paper is the characterization of the space of signatures which enable this property. We show that selecting the optimal signature in this space is NP-complete, and based on insights from the characterization of the space, we propose two novel filters which help to prune the candidates further before verification. In addition, we introduce a simple optimization to the calculation of the maximum matching metric itself based on the triangle inequality. Compared to related approaches, SILKMOTH is much more general, handling a larger space of similarity functions and relatedness metrics, and is an order of magnitude more efficient on real datasets.

Table 1: Two related datasets.

| Location |
| --- |
| 77 Mass Ave Boston MA |
| 5th St 02115 Seattle WA |
| 77 5th St Chicago IL |

| Address |
| --- |
| 77 Massachusetts Avenue Boston MA |
| Fifth Street Seattle MA 02115 |
| 77 Fifth Street Chicago IL |
| One Kendall Square Cambridge MA |

## 1. INTRODUCTION

Data analysts increasingly need to reason about data from a variety of different sources, which may or may not be semantically related. Given two datasets, an analyst will frequently want to understand how similar two data sets are, e.g., to find copying relationships between data, or determine if two datasets are joinable on a particular column, or to perform schema integration on two datasets. In these applications, a natural goal is to have a measure of *set relatedness*, i.e., how similar two sets are, or how much of one set is contained in another set.

A popular metric for set relatedness is to model the two sets as the two sides of a bipartite graph (each element assuming a vertex), use a similarity function to measure the similarity between elements, and measure the overall relatedness score as the average of edge weights in the maximum bipartite matching [11] (with the weight of each edge determined by the similarity function). Unlike other metrics that require the elements of the sets to match exactly, this metric is robust to small dissimilarities between elements. For example in Table 1, although both *Location* and *Address* likely refer to the same entities, none of the elements exactly match.

In real-world situations with dirty data, the robustness of the metric offers a big advantage, and there have been many recent applications which utilize the maximum bipartite matching metric. For example, in schema matching, the metric can be used to measure the similarity of values in columns or names of attributes [12,20,23]. Similarly, in string matching [25], given two columns of strings, each string can be tokenized on whitespace and the metric can be used to find matching pairs of strings or assess the overall similarity of columns. Finally, the maximum bipartite matching metric has been used to find approximate inclusion dependencies [3,7], where two columns are considered joinable if one column approximately contains another.

**Challenges.** Despite the widespread usage and importance of set relatedness, there has been little work on efficiently discovering pairs of related sets from a collection of sets. In particular, determining the maximum bipartite matching of two sets takes $O(n^3)$ time where $n$ is the number of elements in the sets [10,11], and a naive implementation for finding related sets would require the comparison between every pair of sets, leading to $m^2$ time where

---

$m$ is the number of sets. Given that real datasets can have many thousands of elements (e.g., values per column) and hundreds or thousands of sets (e.g., columns per dataset), a total runtime of $O(n^3 m^2)$ can be quite costly.

**Approach.** We propose SILKMOTH[1], a general-purpose system capable of rapidly discovering related sets with the maximum bipartite matching metric. We consider two notions of set relatedness: (1) SET-SIMILARITY which checks whether two sets are approximately equivalent, and (2) SET-CONTAINMENT which checks whether one set is approximately a superset of the other. For the similarity function used to measure closeness between elements, we support both (a) character-based edit similarity and (b) token-based Jaccard similarity, with optional support for imposing a minimum similarity threshold. In addition, SILKMOTH has two modes of operation: (i) Discovery mode, where we search for all pairs of related sets within a dataset. (ii) Search mode, where, given a reference set, we search for all related sets in the dataset. SILKMOTH is guaranteed to produce the exact same output as the naive method, unlike approximate methods [9].

Internally, SILKMOTH operates by creating signatures for each set and using the signatures to select potentially related *candidate* sets from the entire collection. SILKMOTH's signatures have the property that all truly related sets must match the signature; thus although false positives are allowed, there must be no false negatives. In the end, SILKMOTH applies the maximum matching on the remaining candidate sets to eliminate the false positives and verify which of the sets are truly related. For the development of SILKMOTH, we extensively analyzed the problem of creating the *optimal* signature for a set. We were able to derive a full characterization of the space of valid signatures, which ensures no related sets are missed, and show that the optimal signature generation problem is in fact NP-Complete. As a result, SILKMOTH uses heuristic-based algorithms to choose its signatures, which works well in practice, as demonstrated by our experiments. SILKMOTH also makes use of two novel filters after the initial candidate selection to significantly trim down the number of candidate sets. Although the worst-case runtime complexity of SILKMOTH is still $O(n^3 m^2)$, our heuristics ensure that we make far fewer than $m^2$ comparisons, thereby reducing the overall runtime cost by orders of magnitude in practice. SILKMOTH also introduces a novel optimization to the maximum matching problem; the similarity functions (Jaccard similarity and edit similarity) supported by SILKMOTH can take advantage of the triangle inequality to reduce the number of vertices in the bipartite graph for the maximum matching.

There has been prior work that attempts to optimize the set matching problem. However, the current state-of-the-art approach [26] focuses only on the approximate string matching problem. It cannot handle the SET-CONTAINMENT metric, and only supports edit similarity as the similarity function between elements. They also use a signature based approach, but as we show in Section 4, their approach generates a large number of unrelated candidates, which results in significant runtime overheads relative to the optimized approach we propose. In addition, the state-of-the-art approach does not include the additional filtering steps or the optimized maximal matching algorithm. As a result, their approach is orders of magnitude less efficient than SILKMOTH, as we show in our evaluation.

In summary, the main contributions of this paper are:

- Analysis of the signature generation problem, a characterization of the entire space of valid signatures, examination of the optimal signature generation problem, proof of its NP-

Completeness, and multiple heuristic-based algorithms which work well in practice.

- Two novel filtering methods that prune the number of candidates by orders of magnitude, along with a triangle inequality-based optimization to the maximum matching problem.

- Support for both discovery and search modes, various semantics of relatedness, various similarity functions, as well as a minimum threshold for similarity between elements.

- Experiments on real datasets that show SILKMOTH outperforms the state-of-the-art approach by orders of magnitude.

The rest of the paper is organized as follows. Section 2 formally introduces the problems which SILKMOTH solves. Section 3 gives a high level overview of our framework. Section 4 describes how SILKMOTH generates its signatures used to select candidates, and Section 5 discusses our novel filtering mechanisms and our optimization to the maximum matching problem. We discuss our extensions to SILKMOTH to support a minimum similarity threshold in Section 6 and edit similarity in Section 7. We evaluate SILKMOTH in 8, and finally discuss related work in Section 9 and conclude in Section 10.

## 2. PROBLEM FORMULATION

We first provide the notation and definitions necessary to formalize the related set discovery and search problems.

### 2.1 Set Relatedness

We first formalize the notion of what it means for two sets $R$ and $S$ to be related to each other. We assume we are given a similarity function $\phi(x, y)$ which measures the similarity between two elements $x$ and $y$ as a number in $[0, 1]$, with 1 representing complete similarity and 0 representing complete dissimilarity. Although there are many different types of similarity functions (e.g., cosine similarity, Dice similarity [25], Hamming similarity [17]), in this paper, we focus on Jaccard similarity and edit similarity. These are representative of a range of token-based and character-based similarity functions, and the other similarity functions in these two categories can be supported in similar ways by SILKMOTH.

Given two elements $x$ and $y$, Jaccard similarity is defined as $\mathtt{Jac}(x, y) = \frac{|x \cap y|}{|x \cup y|}$ where $x$ and $y$ are two bags of whitespace-delimited words. For example, $\mathtt{Jac}(\{\mathsf{50, Vassar, St, MA}\}, \{\mathsf{50, Vassar, Street, MA}\}) = \frac{3}{5}$. Given two elements $x$ and $y$ as strings, we define the edit similarity as $\mathtt{Eds}(x, y) = 1 - \frac{2LD(x,y)}{|x|+|y|+LD(x,y)}$ in accordance with [19], where $|x|$ and $|y|$ are the lengths of $x$ and $y$ and $LD$ is the Levenshtein distance [21] that returns the minimum number of edit operations (insertion, deletion, and substitution) needed to transform one string to the other. For example, $\mathtt{Eds}$ ("50 Vassar St MA", "50 Vassar Street MA") $= 1 - \frac{2*4}{15+19+4} = \frac{15}{19}$. We also support a slightly different form of edit similarity: $\mathtt{NEds}(x, y) = 1 - \frac{LD(x,y)}{\max(|x|,|y|)}$, and as we see in Section 5.3, the triangle inequality property of $\mathtt{Eds}$ allows us to optimize the evaluation, making it the preferable edit similarity function.

Some applications may want to omit pairs of elements which are low in similarity. For this purpose, a similarity threshold $\alpha$ is provided, and the definition of $\phi$ is adjusted as follows:

$$\phi_\alpha(x, y) = \begin{cases} \phi(x, y) & \text{if } \phi(x, y) \geq \alpha \\ 0 & \text{if } \phi(x, y) < \alpha \end{cases}$$

In the rest of the paper, we use $\phi_\alpha$ and $\phi$ interchangeably for ease of presentation.

---

[1] The silk moth can discover mates up to 7 miles away! [27]

**Table 2:** RELATED SET SEARCH **example: reference set** $R$ **and collection of sets** $\mathcal{S} = \{S_1, S_2, S_3, S_4\}$, **relatedness threshold** $\delta = 0.7$.

| $R$ = Location | $S_1$ | $S_2$ | $S_3$ | $S_4$ |
|---|---|---|---|---|
| $r_1$: ($t_1$=77, $t_2$=Mass, $t_3$=Ave, $t_6$=Boston, $t_8$=MA) | $s_1^1$: ($t_2$ $t_3$ $t_5$ $t_6$ $t_7$) | $s_1^2$: ($t_1$ $t_6$ $t_8$) | $s_1^3$: ($t_1$ $t_2$ $t_3$ $t_4$ $t_6$ $t_8$) | $s_1^4$: ($t_1$ $t_2$ $t_3$ $t_8$) |
| $r_2$: ($t_4$=5th, $t_5$=St, $t_7$=02115, $t_9$=Seattle, $t_{10}$=WA) | $s_2^1$: ($t_1$ $t_2$ $t_4$ $t_5$ $t_6$) | $s_2^2$: ($t_1$ $t_4$ $t_5$ $t_6$ $t_7$) | $s_2^3$: ($t_2$ $t_3$ $t_{11}$ $t_{12}$) | $s_2^4$: ($t_4$ $t_5$ $t_7$ $t_9$ $t_{10}$) |
| $r_3$: ($t_1$=77, $t_4$=5th, $t_5$=St, $t_{11}$=Chicago, $t_{12}$=IL) | $s_3^1$: ($t_1$ $t_2$ $t_3$ $t_4$ $t_7$) | $s_3^2$: ($t_1$ $t_2$ $t_3$ $t_7$ $t_9$) | $s_3^3$: ($t_1$ $t_2$ $t_3$ $t_5$) | $s_3^4$: ($t_1$ $t_4$ $t_5$ $t_6$ $t_9$) |

Given $\phi$, we construct a bipartite graph between $R$ and $S$. Each vertex in the graph represents an element from either $R$ or $S$, and the edges connecting the vertices are weighted using $\phi$. We then find the maximum weighted bipartite matching [11] of this graph, where each vertex in one data set is connected to exactly one vertex in the other, and the sum of the weights of the edges is maximized. The score of this maximum matching is simply the sum of the weights of the edges in this maximum matching alignment. Following the notation from [25,26], we denote this maximum matching score as $|R \mathbin{\widetilde{\cap}}_{\phi_\alpha} S|$. Given $|R \mathbin{\widetilde{\cap}}_{\phi_\alpha} S|$, we can formally define the metrics SET-SIMILARITY and SET-CONTAINMENT as follows:

**DEFINITION 1 (SET-SIMILARITY).** *Given two sets $R$ and $S$ and a similarity function, $\phi_\alpha$, the* SET-SIMILARITY *is defined as:*

$$\texttt{similar}_{\phi_\alpha}(R, S) = \frac{|R \mathbin{\widetilde{\cap}}_{\phi_\alpha} S|}{|R| + |S| - |R \mathbin{\widetilde{\cap}}_{\phi_\alpha} S|}$$

**DEFINITION 2 (SET-CONTAINMENT).** *Given two sets $R$ and $S$ of which $|R| \leq |S|$ and a similarity function, $\phi_\alpha$, the* SET-CONTAINMENT *is defined as:*

$$\texttt{contain}_{\phi_\alpha}(R, S) = \frac{|R \mathbin{\widetilde{\cap}}_{\phi_\alpha} S|}{|R|}$$

**EXAMPLE 1.** *Consider the two columns Address and Location from Table 1. Suppose we use Jaccard similarity to evaluate the closeness between elements and set the similarity threshold to $\alpha = 0.2$. The maximum matching will align the two first tuples, the two second tuples, and the two third tuples with Jaccard similarities of $\frac{1}{3}$, $\frac{1}{5}$, and $\frac{3}{5}$ respectively. All of these are larger than $\alpha$, so $\texttt{contain}_{\phi_\alpha}(Address, Location) = \frac{1/3 + 1/5 + 3/5}{3} \approx 0.42$ and $\texttt{similar}_{\phi_\alpha}(Address, Location) = \frac{1/3 + 1/3 + 3/5}{3 + 4 - (1/3 + 1/3 + 3/5)} \approx 0.22$.*

We denote $\texttt{related}_\phi$ as the general metric for set relatedness which can be either $\texttt{similar}_\phi$ or $\texttt{contain}_\phi$, and two sets are said to be related if and only if $\texttt{related}_\phi(R, S) \geq \delta$ for some given relatedness threshold $\delta$. Note that $\texttt{similar}_\phi(R, S)$ is the fuzzy Jaccard metric (FJACCARD) in [26] when $\phi = \texttt{NEds}$.

### 2.2 Problem Definition

With our definitions for set relatedness, we can now formalize the problems we wish to solve.

**PROBLEM 1 (RELATED SET DISCOVERY).** *Given two collections of sets $\mathcal{R}$ and $\mathcal{S}$, a relatedness threshold $\delta > 0$,[2] and a similarity function $\phi_\alpha$, find all pairs of related sets $\langle R, S \rangle \in \mathcal{R} \times \mathcal{S}$ such that $\texttt{related}_{\phi_\alpha}(R, S) \geq \delta$.*

**PROBLEM 2 (RELATED SET SEARCH).** *Given a reference set $R$, a collection of sets $\mathcal{S}$, a relatedness threshold $\delta > 0$, and a similarity function $\phi_\alpha$, find all sets $S$ in $\mathcal{S}$ related to the reference set $R$ such that $\texttt{related}_{\phi_\alpha}(R, S) \geq \delta$.*

The related set discovery and search problems are closely linked, so SILKMOTH employs a unified framework to solve both problems.

**Running Example.** To aid the reader in understanding the technical details of SILKMOTH, we present an example in Table 2, which we refer to in other sections. In the table, $R$ is the reference set, which for instance could correspond to the **Location** in

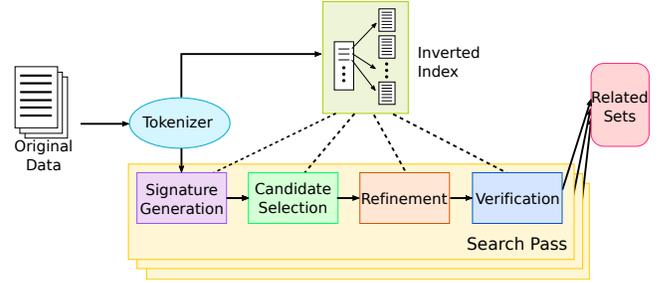

**Figure 1:** SILKMOTH**'s Framework**

Table 1, and $\mathcal{S}$ is the collection of sets from which we search for related sets. Due to space limitations, we use the notation $t_i$ in place of the actual token. For example, we use $t_1$ to denote "77" and $t_2$ to denote "Mass". In addition, each $r_i$ represents the $i^{th}$ element in $R$ and $s_i^j$ represents the $i^{th}$ element in $S_j$. For instance, $r_1$ represents the $1^{st}$ element in **Location** (i.e., $r_1$ ="77 Mass Ave Boston MA"). Furthermore, each element is itself a set of tokens (e.g., $r_1 = \{t_1, t_2, t_3, t_6, t_8\}$ where $t_1$="77", $t_2$="Mass", $t_3$="Ave", $t_6$="Boston", and $t_8$="MA"). For the purpose of our example, there are a total of 12 unique tokens in the dataset, and the tokens are subscripted in decreasing order of frequency (e.g., $t_1$ appears more times than $t_{12}$).

**EXAMPLE 2.** *Using the running example in Table 2, consider the reference set $R$ and the collection of sets $\mathcal{S}$. Suppose $\phi = \texttt{Jac}$, $\alpha = 0$, and the* SET-CONTAINMENT *threshold $\delta = 0.7$. The related set search should return only $S_4$ since $\texttt{similar}_\phi(R, S_4) \approx 0.743 > 0.7$ where $r_1, r_2, r_3$ respectively align with $s_1^4, s_2^4, s_3^4$ and $|R \mathbin{\widetilde{\cap}}_\phi S_4| = \texttt{Jac}(r_1, s_1^4) + \texttt{Jac}(r_2, s_2^4) + \texttt{Jac}(r_3, s_3^4) = 0.8 + 1 + 0.429 = 2.229$, while the* SET-CONTAINMENT *between $R$ and $S_1, S_2$, and $S_3$ are all less than $\delta$.*

## 3. THE FRAMEWORK

This section presents the unified framework SILKMOTH employs for the related set search and discovery problems. An overview of the framework is given by Figure 1. After the original data is ingested, SILKMOTH first produces the sets that correspond to the user application (e.g., columns for approximate inclusion dependency discovery). SILKMOTH then takes the following steps to find related sets:

**Tokenizer.** For each element in each set, the tokenizer creates an array of tokens to represent that element. These tokens are used later to build the inverted index and generate signatures for the sets. Depending on the similarity function, different types of tokens must be generated. For Jaccard similarity, the tokenizer treats each whitespace-delimited word as a token, and for edit similarity, each token is a *q-gram* (a $q$-length substring of the element[3]; e.g., the 4-grams of the element "50 Vassar St MA" are "50 V", "0 Va", etc.).

**Inverted Index.** SILKMOTH uses the tokens from the tokenizer to create an inverted index $\mathcal{I}$ for the dataset. For each token $t$, the inverted list $\mathcal{I}[t]$ is the list of all $\langle set, element \rangle$ pairs which contain token $t$[4]. Figure 2 on the left shows the inverted index based on the collection of sets $\mathcal{S}$ in Table 2. For example, $t_8$ appears in $s_1^2$, $s_1^3$, and $s_1^4$.

---

[2] When $\delta = 0$, all sets are related to each other, and the problem is trivial.

[3] $q - 1$ special characters are padded at the end.

[4] For efficiency, SILKMOTH deduplicates the sets in the inverted lists and uses unique identity numbers for all references.

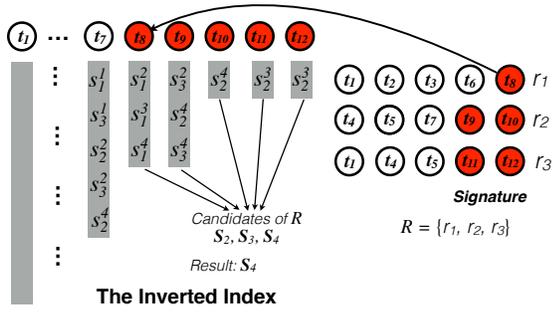

**Figure 2: Tokens and signatures for the example in Table 2.**

**Signature Generation.** SILKMOTH also uses the tokens from the tokenizer to generate a *signature* for each set. Although we delay the full discussion of signature generation until Section 4, in short, SILKMOTH constructs a signature for a set $R$ by selecting the "smallest" set of tokens from $R$ such that if another set $S$ does not share any tokens with $R$'s signature, $R$ and $S$ cannot be related. The tokens in the signature are called *signature tokens*.

**Candidate Selection.** For each signature token $t$ of the set $R$, SILKMOTH accesses the inverted list $\mathcal{I}[t]$ to get all sets which contain $t$. The union of the sets from these lists form the initial candidate sets $\mathcal{C}$. Any set not in $\mathcal{C}$ cannot be related to $R$ due to how the signature is constructed.

**Refinement.** SILKMOTH further prunes $\mathcal{C}$ through the use of more advanced filters which remove non-related sets from $\mathcal{C}$. Full details of these filters are in Section 5, but all sets removed by this step are guaranteed to not be related to $R$.

**Verification.** Finally, SILKMOTH performs the maximum matching between every set in $\mathcal{C}$ with $R$. The sets whose maximum matching scores surpass the relatedness threshold $\delta$ are the final verified related sets to $R$.

For both RELATED SET SEARCH and RELATED SET DISCOVERY, the tokens in $\mathcal{S}$ are used to build the inverted index $\mathcal{I}$. Then for RELATED SET DISCOVERY, for each set $R \in \mathcal{R}$, SILKMOTH selects a signature for $R$, generates candidates based on matches between the signature and $\mathcal{I}$, refines the candidates, and verifies the remaining to obtain the truly related sets. Together, these steps form a related set *search pass*. For RELATED SET SEARCH, only a single search pass needs to be run for the given reference set $R$. Note that for both RELATED SET SEARCH and RELATED SET DISCOVERY, the inverted index $\mathcal{I}$ is only created once in the beginning and used in every search pass thereafter.

Currently, SILKMOTH assumes that both the data and the inverted index can fit in memory. Extensions to external memory and distributed computation are left as future work.

EXAMPLE 3. *Consider the reference set $R$ and the dataset $\mathcal{S}$ in Table 2. Suppose $\phi = $ Jac, $\alpha = 0$, and the SET-SIMILARITY threshold $\delta = 0.7$. We first tokenize all the elements into tokens and then build an inverted index based on the tokens of $\mathcal{S}$. Figure 2 on the left side shows the inverted index. Next we generate a signature for the query $R$ (suppose it is the set of tokens in red in the figure). For each of the five signature tokens, we access its corresponding inverted lists. We union these lists and get three candidates $S_2$, $S_3$, and $S_4$. Finally we verify them by calculating their maximum matching to $R$ and get a result $S_4$ as similar$_\phi(R, S_4) \approx 0.743 > 0.7$.*

Next, we examine the signature generation, refinement, and verification steps in more detail in Sections 4 and 5. Note that the following sections assume that the similarity function $\phi = $ Jac and the similarity threshold $\alpha = 0$. In Section 6, we relax this condition and actually leverage the fact that $\alpha \neq 0$ to make the signatures even more effective. We extend work to edit similarity (i.e., $\phi = $ Eds) in Section 7.

## 4. SIGNATURE GENERATION

A signature used in conjunction with the inverted index allows SILKMOTH to identify the set pairs which are highly likely to be related, without enumerating all set pairs one-by-one. SILKMOTH generates a signature for a set by taking a subset of the tokens in that set. Valid signatures have the property that if set $S$ is related to $R$, then $S$ must share a token in common with $R$'s signature. However, the converse may not necessarily be true; there could be many candidate sets, which contain a token in common with $R$'s signature, but are not actually related to $R$. Thus, the challenge lies in generating a discerning signature which minimizes the number of candidates while still ensuring every related set is a candidate. A perfectly valid, naive signature for $R$ would be to choose all its tokens as the signature. However, this leads to an unnecessary number of candidates. We show how we can generate a better signature in the rest of this section. We formally define what a *valid signature* is in Section 4.1. In Section 4.2 we characterize the space of all valid signatures, which we call the *weighted signature scheme*. Finally, in Section 4.3, we analyze the problem of finding the *optimal* valid signature, show that is is NP-Complete, and provide various approximate heuristic-based algorithms.

### 4.1 Valid Signature

Given a set $R$, we assume that each element $r \in R$ is a set of tokens. We define the set of all tokens in set $R$ as $R^{\mathcal{T}} = \bigcup_{r \in R} r$. Given this, we define a *signature* as follows:

DEFINITION 3 (SIGNATURE). *Given a set $R$, any subset of $R^{\mathcal{T}}$ is a signature of $R$.*

Thus, if we denote $K_R^{\mathcal{T}}$ as a signature for $R$, $K_R^{\mathcal{T}} \subset R^{\mathcal{T}}$. A valid signature has an additional constraint:

DEFINITION 4 (VALID SIGNATURE). *Given a set $R$ and a relatedness threshold $\delta$, a valid signature is a signature $K_R^{\mathcal{T}}$ such that if* related$_\phi(R, S) \geq \delta$ *for any conceivable set $S$, then $S^{\mathcal{T}} \cap K_R^{\mathcal{T}} \neq \varnothing$.*

We define a *signature scheme* to be any set of signatures. Finally, if we are given a signature $K_R^{\mathcal{T}}$ for set $R = \{r_1, r_2, \cdots, r_n\}$, we define the *unflattened signature* as $K_R = \{k_1, k_2, \cdots, k_n\}$ where $k_i = r_i \cap K_R^{\mathcal{T}}$ (i.e., $k_i$ is $r_i$'s set of signature tokens).

EXAMPLE 4. *For the set $R$ in Table 2, we have $R^{\mathcal{T}} = r_1 \cup r_2 \cup r_3 = \{t_1, t_2, \cdots, t_{12}\} = \{77, Mass, \cdots, IL\}$. The subset $K_R^{\mathcal{T}} = \{t_1, t_4, t_5\} = \{77, 5th, St\}$ of $R^{\mathcal{T}}$ is one possible signature of $R$. Its corresponding unflattened signature is $K_R = \{\{t_1\}, \{t_4, t_5\}, \{t_1, t_4, t_5\}\} = \{\{77\}, \{5th, St\}, \{77, 5th, St\}\}$.*

### 4.2 Weighted Signature Scheme

**Maximum Matching Threshold.** When dealing with signatures, rather than directly using the relatedness threshold $\delta$, we instead define a related quantity, the *maximum matching threshold*, denoted as $\theta$. The maximum matching threshold is based on our definitions of similar$_\phi(R, S)$ and contain$_\phi(R, S)$. For contain$_\phi$, we have that $R$ is related to $S$ if contain$_\phi(R, S) = \frac{|R \widetilde{\cap}_\phi S|}{|R|} \geq \delta$. This yields $|R \widetilde{\cap}_\phi S| \geq \delta |R|$, and $R$ is related to $S$ if and only if the maximum matching score between $R$ and $S$ is at least $\delta |R|$. We thus define the maximum matching threshold for contain$_\phi$ as $\theta = \delta |R|$. For similar$_\phi$, $R$ and $S$ are only related if and only if similar$_\phi(R, S) = \frac{|R \widetilde{\cap}_\phi S|}{|R| + |S| - |R \widetilde{\cap}_\phi S|} \geq \delta$. Given that

$|S| \geq |R \widetilde{\cap}_\phi S|,$[5] we have:

$$\delta \leq \frac{|R \widetilde{\cap}_\phi S|}{|R| + |S| - |R \widetilde{\cap}_\phi S|} \leq \frac{|R \widetilde{\cap}_\phi S|}{|R| + |S| - |S|} = \frac{|R \widetilde{\cap}_\phi S|}{|R|}.$$

Thus, $|R \widetilde{\cap}_\phi S| \geq \delta |R|$. Therefore, the maximum matching threshold for `similar`$_\phi$ is also $\theta = \delta |R|$. The maximum matching threshold is particularly useful because it only depends on the set $R$ for which we are building the signature. As we will see in the following, this allows SILKMOTH to build a single signature for $R$ which can be used and compared against all sets in $\mathcal{S}$.

**Unweighted Signature Scheme.** We first introduce the unweighted signature scheme similar to the current state-of-the-art [25]. This work reasoned that for the maximum matching score to be at least $\theta$, there must be at least $c = \lceil \theta \rceil$ pairs of elements which have a similarity score larger than 0, so at least $c$ elements must share a token in common. In other words for two sets to be related, the two sets must share at least $c$ tokens in common. Based on this observation, the state-of-the-art approach is to remove $c - 1$ tokens in $R^\mathcal{T}$ (treating $R^\mathcal{T}$ as a multiset) and use the union of the rest of tokens as the signature $K_R^\mathcal{T}$. If $R$ and $S$ are related, at least one of the $c$ tokens $R$ and $S$ share must remain in $K_R^\mathcal{T}$, and $K_R^\mathcal{T} \cap S^\mathcal{T} \neq \varnothing$. Thus $K_R^\mathcal{T}$ must be a valid signature.

EXAMPLE 5. *Consider the example in Table 2, with $\theta = \delta|R| = 0.7 * 3 = 2.1$. We can remove $c - 1 = \lceil \theta \rceil - 1 = 2$ tokens and use the union of the remaining tokens as a valid signature. Suppose we remove $t_{11}$ and $t_{12}$ from $r_3$, the signature $K_R^\mathcal{T} = \{t_1, t_2, t_3, t_6, t_8\} \cup \{t_4, t_5, t_7, t_9, t_{10}\} \cup \{t_1, t_4, t_5\} = \{t_1, t_2, \cdots, t_{10}\} = \{77, Mass, \cdots, WA\}$ is valid.*

In short, the *unweighted signature scheme* for set $R = \{r_1, \cdots, r_n\}$ contains all signatures $K_R^\mathcal{T}$ with a corresponding unflattened signature $K_R = \{k_1, k_2, \cdots, k_n\}$ such that $\sum_{i=1}^{n} |r_i| - |k_i| \leq c - 1$.

**Drawbacks of Unweighted Signature Scheme.** While all signatures in the unweighted signature scheme are valid signatures, it captures only a small subset of the whole space of the valid signatures and yields a large number of spurious candidates. By only requiring $c = \lceil \theta \rceil$ shared tokens, the unweighted signature scheme effectively estimates an upper bound of 1 for the similarity score between an element $r \in R$ and any element $s$ sharing a token with $r$. This results in an upper bound of $c$ for the total maximum matching score. However, in many cases, this is a severe overestimation of the upper bound. For example, consider $r_3$ and $s_1^1$ in Table 2, which share token $t_5$. Their similarity score is $\frac{1}{5+5-1} \ll 1$.

**Weighted Signature Scheme.** The *weighted signature scheme* captures all signatures in the unweighted scheme but also considers signatures based on a tighter upper bound for the similarity score, which in turn leads to a tighter upper bound for the total maximum matching score. Given that the Jaccard similarity between any elements $r$ and $s$ is calculated as $\frac{|r \cap s|}{|r \cup s|}$, for any element $r$, consider an element $s$ which shares only one token with $r$; it must have a similarity score of $\frac{1}{|r| + |s| - 1} \leq \frac{1}{|r|}$. Now consider an element $s$ which shares $x$ tokens with $r$; its similarity score is $\frac{x}{|r| + |s| - x} \leq \frac{x}{|r|}$. This upper bound $\frac{x}{|r|}$ attributes a *weight* of $\frac{1}{|r|}$ to each token in $r$, representing an upper bound to each token's contribution to the overall maximum matching score. In contrast, the unweighted signature scheme is equivalent to assuming each token's contribution to the overall maximum matching score is 1. This finer granularity allows the weighted signature scheme to create smaller signatures than the unweighted signature scheme and reduce spurious candidates; in fact,

Section 8.2 shows that the weighted signature scheme provides an improvement of up to $7.7\times$ in performance compared to the unweighted signature scheme. We next extend the idea to formally define the weighted signature scheme.

DEFINITION 5 (WEIGHTED SIGNATURE SCHEME). *Given the set $R = \{r_1, \cdots, r_n\}$ and a related threshold $\delta$, the weighted signature scheme is the family of all signatures $K_R^\mathcal{T}$ which have unflattened signature $K_R = \{k_1, \cdots, k_n\}$ satisfy $\sum_{i=1}^{n} \frac{|r_i| - |k_i|}{|r_i|} < \delta|R|$.*

EXAMPLE 6. *Consider the set $R$ in Table 2 and suppose $\delta = 0.7$, yielding $\theta = |R|\delta = 3*0.7 = 2.1$. $K_R^\mathcal{T} = \{t_8, t_9, t_{10}, t_{11}, t_{12}\} = \{MA, Seattle, WA, Chicago, IL\}$ is a valid signature in the weighted signature scheme with unflattened signature $K_R = \{\{t_8\}, \{t_9, t_{10}\}, \{t_{11}, t_{12}\}\} = \{\{MA\}, \{Seattle, WA\}, \{Chicago, IL\}\}$ as shown in Figure 2. Note $\sum_{i=1}^{n} \frac{|r_i| - |k_i|}{|r_i|} = \frac{5-1}{5} + \frac{5-2}{5} + \frac{5-2}{5} = 2 < \theta$.*

There are two interesting aspects to the weighted signature scheme. First, all signatures in the weighted signature scheme are valid signatures (Lemma 1). Second, there are no valid signatures outside the weighted signature scheme (Lemma 2). So, we can state:

> THEOREM 1. *The weighted signature scheme is exactly the set of all valid signatures for a given set and relatedness threshold.*

LEMMA 1. *All signatures in the weighted signature scheme are valid.*

PROOF. Given a set $R = \{r_1, r_2, \cdots, r_n\}$, consider a signature $K_R^\mathcal{T}$ in the weighted signature scheme and its unflattened signature $K_R = \{k_1, k_2, \cdots, k_n\}$. For any set $S$ which satisfies $S^\mathcal{T} \cap K_R^\mathcal{T} = \varnothing$, we derive an upper bound for the maximum matching score between $R$ and $S$. For any element $s$ in $S$, the similarity between $r_i$ and $s$ is $\frac{|r_i \cap s|}{|r_i \cup s|}$. First, it is trivially true that $|r_i \cup s| \geq |r_i|$ for all $r_i$ and $s$. Second, given that $S^\mathcal{T} \cap K_R^\mathcal{T} = \varnothing$, $k_i \cap s = \varnothing$ for all $k_i$. So, $|r_i \cap s| = |(r_i \setminus k_i) \cap s| + |k_i \cap s| \leq |r_i| - |k_i|$. We now have the upper bound: $\frac{|r_i \cap s|}{|r_i \cup s|} \leq \frac{|r_i| - |k_i|}{|r_i|}$. This is true for all $r_i$, so the upper bound for the overall maximum matching score is $\sum_{i=1}^{n} \frac{|r_i| - |k_i|}{|r_i|}$. Thus, if $\sum_{i=1}^{n} \frac{|r_i| - |k_i|}{|r_i|} < \theta$, we have shown that for all sets $S$ for which $K_R^\mathcal{T} \cap S^\mathcal{T} = \varnothing$, $\text{related}_\phi(R, S) \leq \sum_{i=1}^{n} \frac{|r_i| - |k_i|}{|r_i|} < \theta$, so they cannot be related. □

LEMMA 2. *There are no valid signatures outside the weighted signature scheme.*

PROOF. Assume to the contrary that there is a valid signature $K_R^\mathcal{T}$ and it is not within the weighted signature scheme, i.e., $\sum_{i=1}^{n} \frac{|r_i| - |k_i|}{|r_i|} \geq \theta$. We can construct a set $S = \{s_1, s_2, \cdots, s_n\}$ where $s_i = r_i \setminus k_i$. Obviously $S^\mathcal{T} \cap K_R^\mathcal{T} = \varnothing$. However, by aligning $r_i$ with $s_i$, we have a lower bound for the maximum matching score between $R$ and $S$ as $\sum_{i=1}^{n} \frac{|r_i \cap s_i|}{|r_i \cup s_i|} = \sum_{i=1}^{n} \frac{|r_i| - |k_i|}{|r_i|} \geq \theta$, which means $R$ and $S$ are related. This is a contradiction, thus $K_R^\mathcal{T}$ cannot be a valid signature. □

### 4.3 Optimal Signature Selection

Now that we have characterized the entire space of valid signatures, we describe how to select the optimal signature. Our candidate selection step in Section 3 looks up each token in $K_R^\mathcal{T}$ in our inverted index $\mathcal{I}$ and chooses the union of the sets in the inverted lists as candidates. Therefore, it is in our best interest to minimize the size of this union. In this paper, rather than taking the union, which would make the problem even more complex, we use the total length of these inverted lists, which is proportional to the size of the union, as our optimization goal.

---

[5]There are at most $|S|$ edges in the maximum matching between $R$ and $S$, and each edge has a maximum score of at most 1.

PROBLEM 3 (OPTIMAL VALID SIGNATURE SELECTION). *Given a set $R$ and a relatedness threshold $\delta$, find the valid signature $K_R^T$ which minimizes $\sum_{t \in K_R^T} |\mathcal{I}[t]|$.*

Unfortunately, the optimal signature selection problem is NP-Complete. We observe that this problem looks similar to the knapsack problem. Thus, the proof of NP-Completeness follows a similar approach as the proof for the knapsack problem. Specifically, we first reduce the 3-CNF-SAT problem [14] to an inverse-prime subset sum problem which, given a number $s$ and a multi-set $\mathcal{A}$ of numbers, finds a subset of $\mathcal{A}$ whose sum is exactly $s$. Note all the numbers are in the form $\sum_{p \in P_i} 1/p$ where $P_i \subseteq P$, $P = \{p_1, p_2, \cdots, p_l\}$, $p_i$ is the $(i+3)^{th}$ prime (i.e., $p_1 = 7, p_2 = 11, \cdots$), and $l$ is an integer. Then we reduce the inverse-prime subset sum problem to the decision version of our optimal weighted signature selection problem. We leave the formal proof to the appendix.

> THEOREM 2. *The optimal valid signature selection problem is NP-Complete.*

However, since the problem is similar to the knapsack problem, we can employ many of the greedy approximation algorithms used for the knapsack problem. More specifically, given a set $R = \{r_1, r_2, \cdots, r_n\}$, for each token $t$ in $R^T$, we assign it with a value `value` $= \sum_{r_i | t \in r_i} \frac{1}{|r_i|}$ (recall the weight of a token in Section 4.2) and a cost `cost` $= |\mathcal{I}[t]|$. Since, our objective is to minimize the sizes of the inverted lists, we rank all tokens in $R^T$ by the metric $\frac{\texttt{cost}}{\texttt{value}}$ in increasing order and select the tokens one at a time until the condition in Definition 5 is satisfied, which results in a valid signature based on Theorem 1. The selected tokens are the signature tokens of $K_R^T$.

EXAMPLE 7. *Consider Table 2. The inverted index is shown in Figure 2. The* `costs` *(i.e., the inverted list lengths) for the 12 tokens $t_1, \cdots, t_{12}$ in $R^T$ are respectively 9, 8, 7, 6, 6, 6, 5, 3, 3, 1, 1, and 1. The* `values` *are $\frac{2}{5}$ for $t_1, t_4, t_5$ and $\frac{1}{5}$ for the rest of tokens. We rank them based on $\frac{\texttt{cost}}{\texttt{value}}$ in increasing order and select them in sequence. We first select $t_{12}$, resulting in $|k_3| = 1$ while $|k_1| = |k_2| = 0$. Thus, $\sum_{i=1}^{n} \frac{|r_i| - |k_i|}{|r_i|} = 2.8$, which is greater than $\theta = 2.1$. $\theta$, then we select $t_{11}, t_{10}$, and $t_9$. Then, if we select $t_8$, we have $|k_1| = 1$ and $|k_2| = |k_3| = 2$. So, $\sum_{i=1}^{n} \frac{|r_i| - |k_i|}{|r_i|} = 2.0$, which is less than $\theta = 2.1$. Thus we stop at $t_8$ and have a valid signature $K_R^T = \{t_{12}, t_{11}, t_{10}, t_9, t_8\}$.*

**Summary:** *In this section, we described our methodology for creating signatures for sets (assuming $\phi = Jac$ and $\alpha = 0$), and showed that the problem of picking the optimal signature set is NP-complete. SILKMOTH can then solely use these signatures to remove many unrelated sets from consideration, leaving few sets as candidates. The lemmas from this section prove that any sets not found using the signatures cannot be related (i.e., we do not miss any related sets). In the next section, we focus on how to further trim these candidates using several filters.*

## 5. REFINEMENT AND VERIFICATION

Though our signature scheme removes many unrelated sets from consideration, it still produces many spurious candidates. To further prune the candidates for false positives, we add a refinement step which directly compares $R$ with candidate $S$ and rejects sets if certain bounds do not hold. Note that we could not construct these bounds during the signature generation step, since we only had access to $R$ and not $S$. However, once we have pruned the majority of sets, it is more efficient to perform these refinements first before moving on to the computationally expensive maximum

---

**Algorithm 1:** Candidate Selection and Check Filter

**Input:** $R$: reference, $K_R^T$: signature, $\mathcal{I}$: inverted index
**Output:** $\mathcal{C}$: candidates

1   $\mathcal{C} \leftarrow \{\}$;
2   **foreach** $1 \leq i \leq n$ **do**
3     **foreach** $t \in k_i$ **do**
4       **foreach** $\langle S, s \rangle \in \mathcal{I}[t]$ **do**
5         **if** $\phi(r_i, s) \geq \frac{|r_i| - |k_i|}{|r_i|}$ **then**
6           **if** $S \notin \mathcal{C}$ **then** $\mathcal{C}[S] \leftarrow \varnothing$ ;
7           $\mathcal{C}[S] \leftarrow \mathcal{C}[S] \cup r_i$;

8   **return** $\mathcal{C}$

---

matching verification step[6] The refinement step consists of two filters: the check filter (Section 5.1) and the nearest neighbor filter (Section 5.2). We also accelerate the maximum matching verification step through a triangle inequality-based optimization (Section 5.3), which nets $30 - 50\%$ performance benefits, and discuss the overall pseudocode for SILKMOTH (Section 5.4).

### 5.1 Check Filter

In the previous section, we showed that if set $S$ and signature $K_R^T$ (for set $R$) do not have a token in common, $\phi(r_i, s) \leq \frac{|r_i| - |k_i|}{|r_i|}$ is a valid bound for all $s \in S$. From this, we can derive the overall maximum matching bound $\sum_{i=1}^{n} \frac{|r_i| - |k_i|}{|r_i|} < \theta$, allowing us to prune $S$ if no tokens are shared. However, if $S$ and $K_R^T$ do have a token in common—that is if $k_i \cap s \neq \varnothing$ for some $k_i \in K_R$ and for $s \in S$—the bound $\phi(r_i, s) \leq \frac{|r_i| - |k_i|}{|r_i|}$ may no longer hold. On the other hand, now that we have identified the element $s$ which shares a token $k_i$, we can simply go calculate $\phi(r_i, s)$. If it turns out that $\phi(r_i, s) \leq \frac{|r_i| - |k_i|}{|r_i|}$ still holds for all $k_i$ and $s$ which have a shared token, then the overall bound $\sum_{i=1}^{n} \frac{|r_i| - |k_i|}{|r_i|} < \theta$ still holds, thus we can prune $S$ as a candidate even though $S$ shares a token with $K_R^T$. We call this the *check filter*.

Algorithm 1 provides a pseudocode of how SILKMOTH performs candidate selection and applies the check filter. For each token in the signature, the list of (set, element) pairs which match the signature are retrieved. If the set passes the check filter in lines 5-6, the set and matching elements are added to the collection of candidates.

EXAMPLE 8. *Consider the example in Table 2. Suppose $\delta = 0.7$ and we have the valid signature $K_R^T$ with unflattened signature $K_R = \{\{t_8\}, \{t_9, t_{10}\}, \{t_{11}, t_{12}\}\}$. We access the corresponding inverted lists and test the check filter on the three candidates $S_2$, $S_3$, and $S_4$. $S_2$ does not pass the check filter as $Jac(r_1, s_1^2) = 0.6 < \frac{|r_1| - |k_1|}{|r_1|} = 0.8$ and $Jac(r_2, s_3^2) = 0.25 < \frac{|r_2| - |k_2|}{|r_2|} = 0.6$. $S_3$ and $S_4$ pass the check filter as $Jac(r_1, s_3^3) = \frac{5}{6} \geq \frac{|r_1| - |k_1|}{|r_1|} = 0.8$ and $Jac(r_2, s_4^3) = 1 > \frac{|r_2| - |k_2|}{|r_2|} = 0.6$.*

### 5.2 Nearest Neighbor Filter

The nearest neighbor filter comes from the simple idea that the maximum matching score between $R$ and $S$ is at most the sum of the similarities of each element in $R$ to its most similar element (nearest neighbor) in $S$:

$$|R \widetilde{\cap}_\phi S| \leq \sum_{r \in R} \max_{s \in S} \phi(r, s)$$

In other words, for each element $r \in R$, there is a nearest neighbor $s \in S$ which maximizes $\phi(r, s)$, and the sum of the similarities to the nearest neighbors must be greater than or equal to the maximum matching score. So, if there is a candidate $S$ whose nearest neighbor similarity sum is less than $\theta$, we can prune that candidate. While this filter is much more powerful than the check filter, calculating nearest neighbors can be computationally expensive (though not as expensive as maximum matching). We augment the filter with several techniques to make it more efficient.

---

[6] We also perform a size check for SET-SIMILARITY to compare only similar size sets.

**Efficient Nearest Neighbor Search.** The most time consuming component of the nearest neighbor filter is unsurprisingly the search for the nearest neighbor of each element. To address this, we adapt the technique presented in [28] for our problem. To find the nearest neighbor of element $r \in R$ in the elements of $S$, we traverse through each token $t \in r$, and for each token $t$, we fetch all elements $s$ in $S$ which contain token $t$ using the inverted list $\mathcal{I}[t]$[7]. We calculate the similarities between $r$ and each $s$ we retrieve; the $s$ with the largest similarity score is considered the nearest neighbor. Note that due to the way our tokens are defined, $r$ and $s$ must share a token to have a non-zero similarity score. Thus, it is sufficient to search through the inverted index.

**Computation Reuse.** Recall that in the check filter, we calculated the actual similarities between $r$ and all elements in $s \in S$ which contain a signature token of $r$. The largest of these similarities which exceed $\frac{|r|-|k|}{|r|}$ must be the nearest neighbor similarity since all other elements which do not contain a signature token of $r$ cannot surpass the $\frac{|r|-|k|}{|r|}$ bound, as explained above. Thus, we can reuse the computation from the check filter.

**Early Termination.** For elements in $r \in R$ whose signature tokens appear in candidate $S$, their nearest neighbor similarities are not guaranteed to be bounded by $\phi(r, s) \leq \frac{|r|-|k|}{|r|}$. So, we must perform the nearest neighbor search for all these elements either by reusing computation from the check filter or by actually searching for the nearest neighbor. However, for elements in $r' \in R$ whose signature tokens do not appear in $S$, the bound $\frac{|r'|-|k'|}{|r'|}$ still holds for all $s \in S$. To perform early termination, we first deduce a total estimate by summing the nearest neighbor similarities for matching elements $r$ and using estimates $\frac{|r'|-|k'|}{|r'|}$ for non-matching elements $r'$. Then, we iterate through every $r'$ and update the total estimate with the nearest neighbor similarity of $r'$. If the total estimate ever falls below $\theta$, the candidate $S$ cannot be related since the actual nearest neighbor similarities of $r'$ cannot exceed $\frac{|r'|-|k'|}{|r'|}$, so the overall nearest neighbor score cannot exceed $\theta$.

Algorithm 2 provides the pseudocode of how SILKMOTH applies the nearest neighbor filter. The input candidates $\mathcal{C}$ is assumed to have the same format as the output of the check filter, and we abstract away the nearest neighbor search as NNSearch in the pseudocode. For each candidate $S$, a total estimate is constructed in line 3, the nearest neighbor similarities for matching candidates are computed in lines 4-5, and nearest neighbor similarities for non-matching candidates along with early termination are lines 6-9.

EXAMPLE 9. *Consider the example in Table 2 and suppose* $\delta = 0.7$. *Consider the candidate* $S_3$. *Since* $r_1$ *shares signature tokens with* $s_1^3$. *The nearest neighbor of* $r_1$ *is* $s_1^3$ *with a similarity score* $\frac{5}{6}$. *For* $r_2$, *none of the elements in* $S_3$ *contain any of the signature tokens of* $r_2$, *so we can safely give* $r_2$ *a similarity score upper bound of* $\frac{|r_2|-|k_2|}{|r_2|} = 0.6$. *For* $r_3$, *though* $s_2^3$ *contains the signature tokens of* $r_3$, *it does not pass the check filter, and we can also give* $r_3$ *an upper bound of* $\frac{|r_3|-|k_3|}{|r_3|} = 0.6$. *Then, we calculate the nearest*

---

**Algorithm 2:** Nearest Neighbor Filter
***
**Input:** $R$: reference, $\delta$: threshold, $\mathcal{I}$: inverted index, $\mathcal{C}$: candidates

**Output:** $\mathcal{C}'$: refined candidates

**1** $\mathcal{C}' \leftarrow \varnothing$;

**2 foreach** $S \in \mathcal{C}$ **do**

**3**     total $\leftarrow \sum_{i=1}^{n} \frac{|r_i|-|k_i|}{|r_i|}$;

**4**     **foreach** $r \in \mathcal{C}[S]$ **do**

**5**        $\lfloor$ total $\leftarrow$ total $+$ NNSearch$(r, S, \mathcal{I}) - \frac{|r|-|k|}{|r|}$;

**6**     **foreach** $r \in (R \setminus \mathcal{C}[S])$ **do**

**7**        total $\leftarrow$ total $+$ NNSearch$(r, S, \mathcal{I}) - \frac{|r|-|k|}{|r|}$;

**8**        **if** *total* $< \theta$ **then**

**9**           $\lfloor$ goto 3;

**10**     $\mathcal{C}' \leftarrow \mathcal{C}' \cup S$;

**11 return** $\mathcal{C}'$
***

*neighbor of* $r_2$ *and find* $s_3^3$ *with a similarity* $0.125$. *Thus the similarity score upper bound of* $r_2$ *is updated to* $0.125$. *Since the total estimate is now* $\frac{5}{6} + 0.6 + 0.125 < \theta = 2.1$, *we can early terminate the nearest neighbor filter and prune the candidate* $S_3$.

### 5.3 Reduction-Based Verification

In this section, we present an optimization to the maximum matching verification, the most computationally expensive component in SILKMOTH's framework. Let $\psi(r, s) = 1 - \phi(r, s)$ be the dual *distance function* of the similarity function $\phi$. We observe that if the distance function $\psi$ is in the metric space, (i.e., satisfies the triangle inequality), the identical elements in the two sets must appear in the maximum matching. Consider a pair of identical elements $r$ and $s$. For any other elements $r'$ and $s'$, we have:

$$\phi(r, s') + \phi(r', s) = 1 - \psi(r, s') + 1 - \psi(r', s) \quad (1)$$
$$= 2 - \psi(r, s') - \psi(r', r) \quad (2)$$
$$\leq 2 - \psi(r', s') \quad (3)$$
$$= 1 + 1 - \psi(r', s') \quad (4)$$
$$= \phi(r, s) + \phi(r', s'). \quad (5)$$

In (1), we convert the similarity functions to their dual distance functions. We use the fact that $r = s$ in (2) and apply the triangle inequality in (3). From (4) to (5), we use the fact that $\phi(r, s) = 1$ since $r = s$. In (5), we see that the alignment connecting $r$ to $s$ is better than the alignment which connects $r$ to $s'$ and $r'$ to $s$, for all $r'$ and $s'$. Thus, $r$ and $s$ must exist in the maximum matching.

To apply this reduction, we remove all identical elements from $R$ and $S$ and apply the maximum matching on the resulting sets. After the maximum matching, we add the number of identical elements to the maximum matching score to obtain the final maximum matching score.

### 5.4 Putting It All Together

Algorithm 3 presents the overall pseudocode for SILKMOTH. The inputs are the collections of sets $\mathcal{R}, \mathcal{S}$, the relatedness threshold $\delta$, and the similarity function $\phi$. The output is the collection of all related set pairs. Initially, SILKMOTH builds the inverted index. Then for each set $R$ in $\mathcal{R}$, SILKMOTH generates the signature of $R$ according to Section 4, selects the initial candidates and runs the check filter using the code from Algorithm 5.1, applies the nearest neighbor filter according to Algorithm 5.2, and finally performs verification to confirm whether a candidate is related.

---

**Algorithm 3: SILKMOTH**

**Input:** $\mathcal{R}$: collection, $\mathcal{S}$: collection, $\delta$: threshold, $\phi$: similarity function

**Output:** $\mathcal{A} = \{\langle R, S\rangle \mid \text{related}_\phi(R, S) \geq \delta\}$

1   $\mathcal{A} \leftarrow \varnothing$;
2   $\mathcal{I} \leftarrow \text{BuildInvertedIndex}(\mathcal{S})$;
3   **foreach** $R \in \mathcal{R}$ **do**
4      $K_R^\mathcal{T} \leftarrow \text{SigGen}(R, \delta, \phi, \mathcal{I})$;
5      $\mathcal{C} \leftarrow \text{CandSelCheck}(R, K_R^\mathcal{T}, \mathcal{I})$;
6      $\mathcal{C}' \leftarrow \text{NearestNeighborFilter}(R, \delta, \mathcal{I}, \mathcal{C})$;
7      **foreach** $S \in \mathcal{C}'$ **do**
8         **if** $\text{Verify}(R, S)$ **then**
9            $\mathcal{A} \leftarrow \mathcal{A} \cup \langle R, S\rangle$;

10   **return** $\mathcal{A}$

**Summary:** *In this section, we showed how we can further reduce the number of candidate set pairs by (1) checking that the elements which matched the signatures surpass their estimated similarities, and (2) evaluating the similarities of elements and their nearest neighbors and making sure this overestimate is greater than $\theta$. Once the filters have pruned the set pairs into the final candidates, SILKMOTH then verifies which of these final candidates are truly related by actually computing the bipartite maximum matching for every remaining candidate set pair. We also described an optimization to the actual bipartite maximum matching computation using the triangle inequality. So far, we have assumed that $\alpha = 0$ ($\phi = Jac$), but in the next section, we relax this condition and show how all our techniques extend to the case in which $\alpha \neq 0$ ($\phi = Jac$), with the exception of the verification optimization.*

## 6. SIMILARITY THRESHOLD

The previous sections all assumed that the similarity threshold $\alpha = 0$. In this section, we extend our techniques to support and leverage the case when $\alpha \neq 0$. First, we must extend the concept of a valid signature to what we call $\alpha$-*valid signature*.

DEFINITION 6 ($\alpha$-VALID SIGNATURE). *Given set $R$, relatedness threshold $\delta$, and similarity threshold $\alpha$, an $\alpha$-valid signature is a signature $K_R^\mathcal{T}$ such that if $\text{related}_{\phi_\alpha}(R, S) \geq \delta$ for some set $S$, then $S^\mathcal{T} \cap K_R^\mathcal{T} \neq \varnothing$.*

Note that 0-valid signatures are the same as the valid signatures we defined in Section 4.1, thus we use them interchangeably. The aim for the rest of this section is to find the best $\alpha$-valid signature for set $R$.

### 6.1 Sim-Thresh Signature Scheme

The similarity threshold $\alpha$ gives us another perspective on how to generate signatures. The weighted signature scheme focuses on selecting signature tokens from the entire set of tokens $R^\mathcal{T}$ for set $R$ based on the maximum matching threshold. Instead, with the *sim-thresh signature scheme*, we can consider selecting signature tokens from individual elements $r \in R$ such that any matching element $s \in S$ must contain one of the signature tokens, otherwise the similarity score would be less than the similarity threshold $\alpha$ and converted to 0. Thus, if no elements in $S$ match any of the signature tokens from any of the elements $r \in R$, the maximum matching score must be less than $\theta$, since the similarity score for every $r \in R$ is 0.

More formally, recall that the similarity score of two elements $r \in R$ and $s \in S$ is $\frac{|r \cap s|}{|r \cup s|} \leq \frac{|r \cap s|}{|r|}$. We deduce how many signature tokens we must pick from $r$ such that this bound is less than $\alpha$ if $s$ does not contain any of those signature tokens. So, $\frac{|r \cap s|}{|r|} < \alpha$

yields $|r \cap s| < \alpha|r|$, and if we select signature tokens $m$ from $r$ where $m \cap s = \varnothing$:

$$|r \cap s| = |(r \setminus m) \cap s| + |m \cap s| \leq |r \setminus m| \leq |r| - |m| < \alpha|r|$$

then we see that we must pick at least $|m|$ signature tokens where $|m| > (1 - \alpha)|r|$. Since $|m|$ is an integer, we require $|m| \geq \lfloor (1 - \alpha)|r| \rfloor + 1$. We can now formally define the sim-thresh signature scheme:

DEFINITION 7 (SIM-THRESH SIGNATURE SCHEME). *Given a set $R = \{r_1, r_2, \cdots, r_n\}$ and a similarity threshold $\alpha$, the sim-thresh signature scheme is the family of all signatures $M_R^\mathcal{T}$ which unflattened signature $M_R = \{m_1, m_2, \cdots, m_n\}$ that satisfies $|m_i| \geq \lfloor (1 - \alpha)|r_i| \rfloor + 1$ for all $1 \leq i \leq n$.*

Note that all signatures in the sim-thresh scheme must be $\alpha$-valid by construction.

EXAMPLE 10. *Consider the example in Table 2 and suppose $\alpha = 0.7$. The signature $M_R^\mathcal{T} = \{t_6, t_8, t_9, t_{10}, t_{11}, t_{12}\}$ is within the sim-thresh signature scheme. This is because $M_R = \{\{t_6, t_8\}, \{t_9, t_{10}\}, \{t_{11}, t_{12}\}\}$ and for all $1 \leq i \leq 3$, $|m_i| = 2 \geq \lfloor (1 - \alpha)|r_i| \rfloor + 1 = \lfloor (1 - 0.7) * 5 \rfloor + 1 = 2$ holds.*

### 6.2 Combined Signature Scheme

The sim-thresh signature scheme, which depends solely on the similarity threshold $\alpha$, can be considered orthogonal to the unweighted and weighted signature schemes, which depend on the relatedness threshold $\delta$. Thus, we can combine and use these signature schemes simultaneously.

Suppose $K_R^\mathcal{T}$ is a signature in the weighted signature scheme, $M_R^\mathcal{T}$ is a signature in the sim-thresh signature scheme, and the unflattened signatures are defined as usual: $k_i = K_R^\mathcal{T} \cap r_i$ and $m_i = M_R^\mathcal{T} \cap r_i$. For each element $r_i$, we can choose the set of signature tokens as either $k_i$ or $m_i$. If some element $s$ does not contain any of the signature tokens for $r_i$, it must be that either $s \cap k_i = \varnothing$ and $\phi_\alpha(s, r_i) \leq \frac{|r_i| - |k_i|}{|r_i|}$, or $s \cap m_i = \varnothing$ and $\phi(s, r_i) < \alpha$ and $\phi_\alpha(s, r_i) = 0$. In both cases $\phi_\alpha(s, r_i) \leq \frac{|r_i| - |k_i|}{|r_i|}$. Thus, the upper bound on the maximum matching score is preserved. Based on this idea, we propose the *combined signature scheme*.

DEFINITION 8 (COMBINED SIGNATURE SCHEME). *Given the set $R = \{r_1, r_2, \cdots, r_n\}$, a relatedness threshold $\delta$, and a similarity threshold $\alpha$, the combined signature scheme is the family of all signatures $L_R^\mathcal{T}$ which have unflattened signature $L_R = \{l_1, l_2, \cdots, l_n\}$ that satisfies $\exists K_R^\mathcal{T}, M_R^\mathcal{T}$ from the weighted and sim-thresh signature schemes respectively such that either $l_i \supseteq k_i$ or $l_i \supseteq m_i$.*

EXAMPLE 11. *Consider the example in Table 2 and suppose $\alpha = \delta = 0.7$. The signature $L_R^\mathcal{T} = \{t_5, t_6, \cdots, t_{12}\}$ is within the combined signature scheme. This is because from previous examples, $K_R^\mathcal{T} = \{t_8, t_9, t_{10}, t_{11}, t_{12}\}$ and $M_R^\mathcal{T} = \{t_6, t_8, t_9, t_{10}, t_{11}, t_{12}\}$ are from the weighted and sim-thresh signature schemes and satisfy $l_i \supseteq m_i$ and $l_i \supseteq k_i$. For example, $l_1 = \{t_6, t_8\} \supseteq m_1 = \{t_6, t_8\}$.*

In the same vein, we can combine the sim-thresh signature scheme with the unweighted signature scheme, to create the *combined unweighted signature scheme*. In fact, this more precisely describes the signature scheme proposed by [25]. The combined signature scheme is a superset of the combined unweighted signature scheme. Furthermore, all signatures in the combined weighted signature scheme are $\alpha$-valid:

> THEOREM 3. *All signatures in combined signature scheme are $\alpha$-valid.*

PROOF. Assume to the contrary that signature $L_R^\mathcal{T}$ for set $R$ in the combined signature scheme (with unflattened signature $L_R = $ $\{l_1, l_2, \cdots, l_n\}$) is not an $\alpha$-valid signature. Then, there exists a set $S$ related to $R$, but $S^{\mathcal{T}} \cap L_R^{\mathcal{T}} = \varnothing$. For all $r_i \in R$ and $s \in S$, $S^{\mathcal{T}} \cap L_R^{\mathcal{T}} = \varnothing$ implies $s \cap l_i = \varnothing$. Since $l_i \supseteq k_i$ or $l_i \supseteq m_i$, either $s \cap k_i = \varnothing$ or $s \cap m_i = \varnothing$.

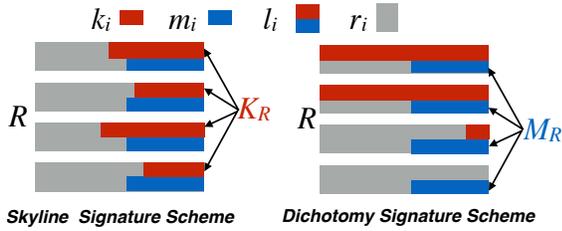

**Figure 3: Skyline and Dichotomy Signature Schemes**

**Case 1:** If $s \cap k_i = \varnothing$, based on the discussion in Section 4.2, we have $\phi_\alpha(r_i, s) \leq \frac{|r_i| - |k_i|}{|r_i|}$.

**Case 2:** If $s \cap m_i = \varnothing$, based on the discussion in Section 6.1, we have $\phi(r_i, s) < \alpha$ and $\phi_\alpha(r_i, s) = 0$.

In both cases, we have $\phi_\alpha(r_i, s) \leq \frac{|r_i| - |k_i|}{|r_i|}$. Thus, the overall maximum matching score between $R$ and $S$ must be no larger than $\sum_{i=1}^{n} \frac{|r_i| - |k_i|}{|r_i|} < \theta$, and $R$ and $S$ cannot be related. This is a contradiction to the original assumption. $\square$

We now discuss how to generate an optimal $\alpha$-valid signature under the combined signature scheme.

**PROBLEM 4 (OPTIMAL $\alpha$-VALID SIGNATURE SELECTION).** *Given a set $R$, relatedness threshold $\delta$, and similarity threshold $\alpha$, find the $\alpha$-valid signature $L_R^{\mathcal{T}}$ under the combined signature scheme that minimizes $\sum_{t \in L_R^{\mathcal{T}}} |\mathcal{I}[t]|$.*

> **THEOREM 4.** *The optimal $\alpha$-valid signature selection problem is NP-Complete.*

As stated in Theorem 6, this problem is NP-Complete when $\alpha = 0$, thus the optimal $\alpha$-valid signature selection problem is also NP-Complete.

### 6.3 The Skyline Signature Scheme

The cardinality of the combined signature scheme is too large, so it is difficult for us to propose an algorithm which effectively solves the optimal $\alpha$-valid signature selection problem. In response, we propose the *skyline signature scheme*, a smaller subset of the combined signature scheme which is guaranteed to contain the optimal signature. The intuition is that instead of allowing $l_i \supseteq k_i$ or $l_i \supseteq m_i$, we impose the more strict condition that $l_i = k_i$ or $l_i = m_i$. Moreover, we add the additional constraint that either $k_i \subseteq m_i$ or $k_i \supseteq m_i$.

**DEFINITION 9 (SKYLINE SIGNATURE SCHEME).** *Given a set $R = \{r_1, r_2, \cdots, r_n\}$, relatedness threshold $\delta$, and similarity threshold $\alpha$, the skyline signature scheme is the family of all signatures $L_R^{\mathcal{T}}$ which have unflattened signature $L_R = \{l_1, l_2, \cdots, l_n\}$ such that $\exists \ K_R^{\mathcal{T}}, M_R^{\mathcal{T}}$ from the weighted and sim-thresh signature schemes respectively where $l_i = k_i \cap m_i$ and either $k_i \subseteq m_i$ or $k_i \supseteq m_i$.*

Figure 3 (left) gives a graphical interpretation of what a skyline signature might look.

> **THEOREM 5.** *The skyline signature scheme contains the optimal $\alpha$-valid signature under the combined signature scheme.*

PROOF. By construction, the skyline signature scheme is a subset of the combined signature scheme. It is straightforward to see that the optimal signature $L_R^{\mathcal{T}}$ from the combined signature must have either $l_i = k_i$ or $l_i = m_i$. Suppose, this optimal signature is derived from $K_R^{\mathcal{T}}$ and $M_R^{\mathcal{T}}$. Our aim is to construct $K_R^{\mathcal{T}'}$ and $M_R^{\mathcal{T}'}$ which show that $L_R^{\mathcal{T}}$ is also a skyline signature. For $\forall i$, we construct the unflattened signature $K_R'$ and $M_R'$ as follows:

$$k_i' = \begin{cases} k_i & \text{if } l_i = k_i \\ k_i \cup m_i & \text{if } l_i = m_i \end{cases} \qquad m_i' = \begin{cases} m_i \cup k_i & \text{if } l_i = k_i \\ m_i & \text{if } l_i = m_i \end{cases}$$

In the newly constructed $K_R^{\mathcal{T}'}$ and $M_R^{\mathcal{T}'}$ it is true by construction that either $k_i' \subseteq m_i'$ or $k_i' \supseteq m_i'$. Furthermore, if we derive $l_i' = k_i' \cap m_i'$, we see that $l_i' = l_i$ for all $i$, thus $L_R^{\mathcal{T}}$ is in the skyline signature scheme. $\square$

With the skyline signature scheme, it is a little easier to see how we can derive an approximate algorithm to find the optimal $\alpha$-valid signature. First, we generate a valid signature $K_R^{\mathcal{T}}$ using the same method as in Section 4.3. Then for the each signature token set $k_i$ in the unflattened signature $K_R$, if $|k_i| < \lfloor (1 - \alpha)|r_i| \rfloor + 1$, we set $l_i = k_i$. Otherwise, we set $l_i$ as the subset of $k_i$ with the $\lfloor (1-\alpha)|r_i| \rfloor + 1$ tokens $t$ that have the minimum $|\mathcal{I}[t]|$. The skyline signature is $L_R^{\mathcal{T}} = \bigcup_{i=1}^{n} l_i$.

**EXAMPLE 12.** *Consider the running example and suppose $\alpha = \delta = 0.7$. We get $K_R^{\mathcal{T}} = \{t_8, t_9, t_{10}, t_{11}, t_{12}\}$ based on the heuristic algorithm in Section 4.3. As $|k_i| \leq \lfloor (1-\alpha)|r_i| \rfloor + 1 = 2$ for all $1 \leq i \leq 3$. We have $l_1 = k_1$, $l_2 = k_2$, and $l_3 = k_3$. Thus $L_R^{\mathcal{T}} = K_R^{\mathcal{T}}$, which is within the skyline signature scheme where $K_R'$ and $M_R^{\mathcal{T}} = \{t_6, t_8, t_9, t_{10}, t_{11}, t_{12}\}$ satisfy the constraints.*

### 6.4 Dichotomy Signature Scheme

We observed that in the skyline signature scheme, whenever $k_i \supseteq m_i$, we can always set $k_i = r_i$, since $l_i = k_i \cap m_i = m_i$ anyway. Since we "added" tokens to $k_i$, this allows us to "remove" tokens from another $k_j \subseteq m_j$, which would further drive down $l_j = k_j \cap m_j = k_j$. We formalize this notion with the *dichotomy signature scheme*.

**DEFINITION 10. (DICHOTOMY SIGNATURE SCHEME)** *Given a set $R = \{r_1, r_2, \cdots, r_n\}$, relatedness threshold $\delta$, and similarity threshold $\alpha$, the dichotomy signature scheme is the family of all signatures $L_R^{\mathcal{T}}$ with unflattened signatures $L_R = \{l_1, l_2, \cdots, l_n\}$ such that $\exists \ K_R^{\mathcal{T}}, M_R^{\mathcal{T}}$ from the weighted and sim-thresh signature schemes respectively for which $l_i = k_i \cap m_i$ and either $k_i = r_i$ or $k_i \subseteq m_i$.*

Figure 3 (right) gives a graphical interpretation of what a dichotomy signature might look like. With the dichotomy signature scheme, we can build a more advanced heuristic algorithm. We add tokens to the signature by the cost/value metric as described in Section 4.3 in ascending order. The first few tokens of an element when $|k_i| < \lfloor (1-\alpha)|r_i| \rfloor + 1$ still have the same cost/value score, once $|k_i| \geq \lfloor (1-\alpha)|r_i| \rfloor + 1$, the cost/value score decreases to 0. Thus, the algorithm is more likely to have entire elements as signature token sets, relying on the sim-thresh signature $m_i$ to cut tokens no smaller than the $\lfloor (1-\alpha)|r_i| \rfloor + 1$ threshold. Note the sim-thresh signature $M_R^{\mathcal{T}}$ is constructed by adding the first $\lfloor (1-\alpha)|r_i| \rfloor + 1$ tokens in $r_i$ with the longest inverted list lengths to $M_R^{\mathcal{T}}$.

**EXAMPLE 13.** *Consider the running example and suppose $\alpha = \delta = 0.7$. We first pick $t_{12}$ whose cost is 1 and value is $\frac{1}{5}$ into the signature. Then we pick $t_{11}$. At this time, we have $k_3 = 2 \geq \lfloor (1-\alpha)|r_3| \rfloor + 1 = 2$. Thus we set the cost/value as 0 for the rest tokens in $r_3$. Then we pick $t_5, t_4, t_1$ which are all cut by the sim-thresh signature $m_1 = \{t_{11}, t_{12}\}$. Now we have $\frac{|r_1| - |k_1|}{|r_1|} + \frac{|r_2| - |k_2|}{|r_2|} + \frac{|r_3| - |k_3|}{|r_3|} = \frac{5-0}{5} + \frac{5-0}{5} + \frac{5-5}{5} = 2 \leq \theta$ and we stop picking tokens into signature. Thus we have $L_R^{\mathcal{T}} = \{t_{11}, t_{12}\}$.*

### 6.5 Refinement and Verification

We also extend the check filter and the nearest neighbor filter to support the similarity threshold. Given a set $R$ and a $\alpha$-valid signature $L_R^T$, instead of checking whether the similarity between $r_i \in R$ and $s_i \in S$ where $l_i \cap s_i \neq \varnothing$ is no smaller than $\frac{|r_i| - |l_i|}{|r_i|}$, the check filter checks if it is no smaller than $\min\left(\alpha, \frac{|r_i| - |l_i|}{|r_i|}\right)$.

For the nearest neighbor filter, for those elements $r_i$ where $|l_i| \geq \lfloor (1 - \alpha)|r| \rfloor + 1$, if another set $S$ does not contain any of the signature tokens in $l_i$, $r_i$ cannot contribute to the maximum matching score between $R$ and $S$, so we estimate the upper bound as $\phi_\alpha(r_i, s) = 0$ for all $s \in S$. We also set the upper bounds for elements $r_i$ with nearest neighbor similarities less than $\alpha$ as 0.

Unfortunately, the reduction-based verification does not work when $\alpha > 0$, as the dual distance function $1 - \phi_\alpha$ does not satisfy the triangle inequality (even though $1 - \phi_{\alpha=0}$ does).

**Summary:** *In this section, we described how our techniques from Sections 4 and 5 can be further specialized for the case $\alpha \neq 0$. We did this by expanding our definition of valid signatures and describing signature schemes which contain $\alpha$-valid signatures. However, the combined signature scheme which contains $\alpha$-valid signatures is too difficult to search over for the optimal signature, so we presented two other signature schemes which are more tractable. We showed that the skyline signature scheme is guaranteed to have the optimal $\alpha$-valid signature, and the dichotomy signature scheme allows us to build an advanced heuristic algorithm to find the optimal $\alpha$-valid signature. Sections 4, 5, and 6 have all been for the case in which $\phi = \text{Jac}$, but in the next section, we extend our work for the case in which $\phi = \text{Eds}$.*

# 7. EDIT SIMILARITY

All previous sections assumed that the similarity function $\phi = \text{Jac}$. We now extend our work to support edit similarity, for both $\phi = \text{Eds}$ and $\phi = \text{NEds}$.

## 7.1 Weighted Signature Scheme

Each token in an element is now represented by a *q-gram* (i.e., a $q$-length substring), and the inverted index is correspondingly constructed using $q$-grams. However, unlike Jaccard similarity, when constructing signatures, only *q-chunks* are used. For a given string $r$, the set of $q$-grams in $r$ correspond to every $q$-length substring in $r$ (e.g., $r[1:q]$, $r[2:q+1]$), whereas the set of $q$-chunks in $r$ correspond to non-overlapping $q$-length substrings which cover the entire string (e.g., $r[1:q]$, $r[q+1:2q]$). Thus, there are $\lceil \frac{|r_i|}{q} \rceil$ $q$-chunks per element $r_i$, where $|r_i|$ now refers to the string length of $r_i$ as opposed to how many tokens it contains.

Just as before, we first give the weighted signature scheme for when $\alpha = 0$. Since the maximum matching score is independent of which similarity function used, the maximum matching threshold $\theta = \delta|R|$ still holds for a given relatedness threshold $\delta$ and a reference set $R$.

We derive an upper bound for the similarity score between an element $r$ which is represented using $q$-chunks and an element $s$ which is represented using $q$-grams; we assume $r$ and $s$ have $x$ matching $q$-chunks/$q$-grams, and have $\lceil \frac{|r_i|}{q} \rceil - x$ mismatching $q$-chunks. The edit distance $LD(r, s) \geq \lceil \frac{|r_i|}{q} \rceil - x$, since at least 1 edit operation is required for each mismatching $q$-chunk. On the other hand, assuming $s$ is the longer string, $|s| - |r| \leq LD(r, s)$[8]. So, we can derive:

$$\text{Eds}(r, s) = 1 - \frac{2LD(r, s)}{|r| + |s| + LD(r, s)} \leq \frac{|r|}{|r| + LD(r, s)}.$$

---

[8]The edit distance between two strings cannot be greater than the difference in their length.

Thus we have $\text{Eds}(r, s) \leq \frac{|r|}{|r| + \lceil |r|/q \rceil - x}$. Moreover,

$$\text{NEds}(r, s) = 1 - \frac{LD(r, s)}{\max(|r|, |s|)} \leq 1 - \frac{LD(r, s)}{|r|} = 1 - \frac{2LD(r, s)}{|r| + |r|}$$

$$\leq 1 - \frac{2LD(r, s)}{|r| + |s| + LD(r, s)} = \text{Eds}(r, s) \leq \frac{|r|}{|r| + \lceil |r|/q \rceil - x}.$$

In both versions of edit similarity, we get the same upper bound. With this bound, we formally define the weighted signature scheme for edit similarity:

**DEFINITION 11.** *Given a set $R = \{r_1, r_2, \cdots, r_n\}$ and a relatedness threshold $\delta$, the weighted signature scheme for edit similarity is the family of all signatures $K_R^T$ which have unflattened signature $K_R = \{k_1, k_2, \cdots, k_n\}$ that satisfy $\sum_{i=1}^{n} \frac{|r_i|}{|r_i| + |k_i|} < \theta$.*

Similar to Jaccard similarity, we can prove the weighted signature scheme is exactly the set of all valid signatures for a given set and relatedness threshold. We can define the optimal valid signature selection problem and use the same heuristic to generate good signatures. It is straightforward to extend the candidate refinement techniques to support the edit similarity as well.

## 7.2 Similarity Threshold

When $\alpha \neq 0$, we can apply similar tricks to generate more discerning signatures. The similarity threshold $\alpha$ gives us another perspective to select valid signatures. Given an element $s$, for any element $s$ with edit similarity no smaller than $\alpha$, we have $\alpha \leq \phi(r, s) \leq \frac{|r|}{|r| + LD(r, s)}$. Thus, $LD(r, s) \leq \frac{1 - \alpha}{\alpha}|r|$ should hold. We can select $\lfloor \frac{1 - \alpha}{\alpha}|r| \rfloor$ $q$-chunks as signature tokens, and any element $s \in S$ which does not share a $q$-gram with the signature tokens must have an edit distance larger than $\lfloor \frac{1 - \alpha}{\alpha}|r| \rfloor$, meaning the similarity is less than $\alpha$. Thus, we can define a sim-thresh signature scheme with $\lfloor \frac{1 - \alpha}{\alpha}|r| \rfloor$ similar to Definition 7. The combined, skyline, and dichotomy signature schemes are all independent of which similarity function is chosen, so all signature generation heuristics still apply.

## 7.3 The Maximum Gram Length

Since the weighted signature scheme contains all the valid signatures, if it is an empty set for a reference set $R$, SILKMOTH cannot generate any valid signature but only compare $R$ with every set $S \in \mathcal{S}$. To avoid the weighted signature scheme to be an empty set, we need the upper bound of $\sum_{i=1}^{n} \frac{|r_i|}{|r_i| + |k_i|}$ to be no smaller than $\theta$. However, the upper bound of $\sum_{i=1}^{n} \frac{|r_i|}{|r_i| + |k_i|}$ depends on $q$; in the weighted signature scheme, when $K_R^T = R^T$, $k_i$ are the set of all $q$-chunks in $r_i$ and $\sum_{i=1}^{n} \frac{|r_i|}{|r_i| + |k_i|} = \sum_{i=1}^{n} \frac{|r_i|}{|r_i| + \lceil |r_i|/q \rceil}$ is maximized. If this is still no smaller than $\theta = n|R|$, the weighted signature scheme is an empty set. As $\sum_{i=1}^{n} \frac{|r_i|}{|r_i| + \lceil |r_i|/q \rceil} \leq \sum_{i=1}^{n} \frac{|r_i|}{|r_i| + |r_i|/q} = n\frac{q}{1+q}$, we need $n\frac{q}{1+q} > \theta = n\delta$. Thus $q < \frac{\delta}{1-\delta}$ should hold to avoid the weighted signature scheme to be empty.

**Summary:** *In this section, we described how the weighted signature scheme changes for $\phi = \text{Eds}$. However, other than this major change, all of our techniques from Sections 5 and 6 still remain applicable to the case in which $\phi = \text{Eds}$.*

# 8. EVALUATION

In this section, we evaluate the runtime performance of SILKMOTH. We show that the techniques introduced in Sections 4, 5, and 6 together greatly cut down the number of candidates which are passed to maximum matching step and provide orders of magnitude improvement in the overall runtime. Figure 4 shows the orders of magnitude in performance gains SILKMOTH achieves with its optimizations for the three applications we experimented with using the default parameters we introduce later. Note that the runtime for inclusion dependency is present but too small to be distinguished. Furthermore, we compared SILKMOTH against the existing FASTJOIN [25] on the approximate string matching application, which FASTJOIN was developed for, and we demonstrate that SILKMOTH outperformed FASTJOIN by up to an order of magnitude.

**Table 3: The Dataset Details**

| Application | Dataset | # Sets | Elems/Set, Tokens/Elem | Problem | Relatedness | $\phi$ | $\theta$ | $\alpha$ |
|---|---|---|---|---|---|---|---|---|
| String Matching | DBLP | 100K | 9, 5 ($q$-gram) | Discovery | SET-SIMILARITY | Eds | (0.7, 0.75, **0.8**, 0.85) | (0.7, 0.75, **0.8**, 0.85) |
| Schema Matching | WEBTABLE | 500K | 3, 11.3 | Discovery | SET-SIMILARITY | Jac | (**0.7**, 0.75, 0.8, 0.85) | (**0.0**, 0.25, 0.5, 0.75) |
| Inclusion Dependency | WEBTABLE | 500K | 22, 2.2 | Search | SET-CONTAINMENT | Jac | (**0.7**, 0.75, 0.8, 0.85) | (0.0, 0.25, **0.5**, 0.75) |

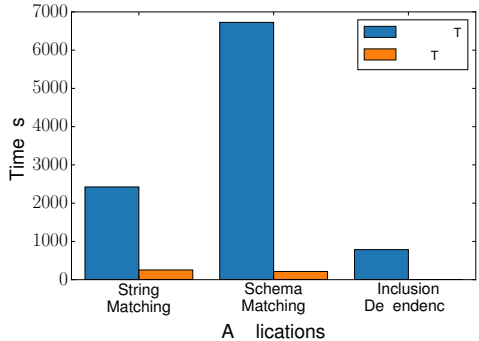

**Figure 4: Overall performance gains of SILKMOTH's optimizations.**

## 8.1 Experimental Setup

We evaluated SILKMOTH on two real-world datasets with three different applications which have been widely used and extensively studied in database systems. The datasets were:

**DBLP**[9]: A bibliography dataset which contains the metadata for computer science publications. We used the publication titles (stripped of punctuation) of 100K randomly chosen publications from 2007 for our experiments.

**WEBTABLE**[10]: A large web data corpus containing millions of HTML tables from the web. We randomly chose 500K tables from the English-language relational subset for our experiments. We treated each web-table as a relational table and only considered columns with non-numerical values.

The three applications were:

**Approximate String Matching:** A RELATED SET DISCOVERY problem under the SET-SIMILARITY metric run on DBLP with each publication title corresponding to a set, a whitespace-delimited word corresponding to an element, and a $q$-gram[11] from the words corresponding to a token. For example, the publication title "`Database System Concepts`" corresponds to a set, which contains three elements "`Database`", "`System`", and "`Concepts`". Suppose $q = 6$, the first element contains 3 tokens or $q$-grams: "`Databa`", "`atabas`", and "`tabase`". We used the same data for both input collections: $\mathcal{R} = \mathcal{S}$, and Eds was the similarity function $\phi$.

**Schema Matching:** A RELATED SET DISCOVERY problem under the SET-SIMILARITY metric run on WEBTABLE with each web-

table schema corresponding to a set, a web-table attribute corresponding to an element, and an attribute value corresponding to a token. We used the same data for both input collections: $\mathcal{R} = \mathcal{S}$, and Jac was the similarity function $\phi$.

**Approximate Inclusion Dependency:** A RELATED SET SEARCH problem under the SET-CONTAINMENT metric run on WEBTABLE with each web-table column corresponding to a set, a column value corresponding to an element, and a whitespace-delimited word from the values corresponding to a token. We randomly chose 500K columns as the dataset $\mathcal{S}$, and from $\mathcal{S}$, we randomly picked 1000 columns to be the references sets $R$. For this application, only columns with more than 4 different values were considered for the random drawings as they were less likely to be categorical variables. Jac was the similarity function.

We varied the relatedness threshold $\delta$ from 0.7 to 0.85, and the similarity threshold $\alpha$ from 0.7 to 0.85 for the string matching application[12] and from 0 to 0.75 for the other two tasks. A summary of the experimental setup is given in Table 3. Default values are highlighted in bold. The number of tokens per element for string matching is based on $\alpha = 0.8$ and $q = 3$. All experiments were completed with a C++-implemented SILKMOTH on a server with 64 Intel(R) Xeon(R) CPU E7-4830 @2.13GHz processors and 256 GB memory. Each experiment measured the overall runtime it took to find related sets, disregarding the time it took to read the data into memory. The time taken to generate the signature was negligible compared to the rest of the runtime, so we do not report those times separately. Every experiment was run 4 times to account for noise. The variance across all runs was less than 1% for all experiments. We measured the memory consumption (dominated mostly by the inverted index and the original dataset) for the string matching, schema matching, and inclusion dependency applications at 24.8MB, 162MB, and 239MB respectively, for datasets of sizes 6.9M, 1GB, and 263MB respectively. For the comparison against FASTJOIN, we received the source of FASTJOIN from its authors and made sure to only consider the related set discovery time.

## 8.2 Signature Generation

*Summary:* SILKMOTH's signature schemes consistently performed better than FASTJOIN's signature scheme for all $\theta$ and $\alpha$ by up to $7.7\times$. Signatures from the dichotomy signature scheme performed best when $\alpha$ was larger, and signatures from the skyline signature scheme performed best when $\alpha$ was smaller.

We first evaluate the signature schemes discussed in this paper. To isolate the effects of the signatures on the overall runtime performance, we disabled all filters for the refinement step and reduction-based optimizations for the verification step for these experiments. The signature schemes we evaluated were based on WEIGHTED: the weighted signature scheme (with no similarity threshold), COMBUNWEIGHTED: the combined unweighted signature scheme which simulates the signature scheme of FASTJOIN but with different token types (FASTJOIN did not use $q$-grams), SKYLINE: the skyline signature scheme from Section 6.3, and DICHOTOMY: the dichotomy signature scheme from Section 6.4. For

---

[9] For each experiment, we chose the maximum possible $q$ given the $\alpha$, based on the correctness constraint $q < \frac{\alpha}{1-\alpha}$ which can be derived from the content in Section 7 (e.g., if $\alpha = 0.85$, then $q = 5$). As long as this constraint is satisfied, the full correctness of our system is guaranteed regardless of the $\alpha$ chosen.

[12] For string matching, it is reasonable that the user wishes to impose a high similarity threshold to avoid extremely different words counting to the overall relatedness [26]. Furthermore, based on the constraint that $q < \frac{\alpha}{1-\alpha}$, if $\alpha = 0.65$, $q = 1$; this led to runtimes which took too long for experiments. FASTJOIN has a similar limitation.

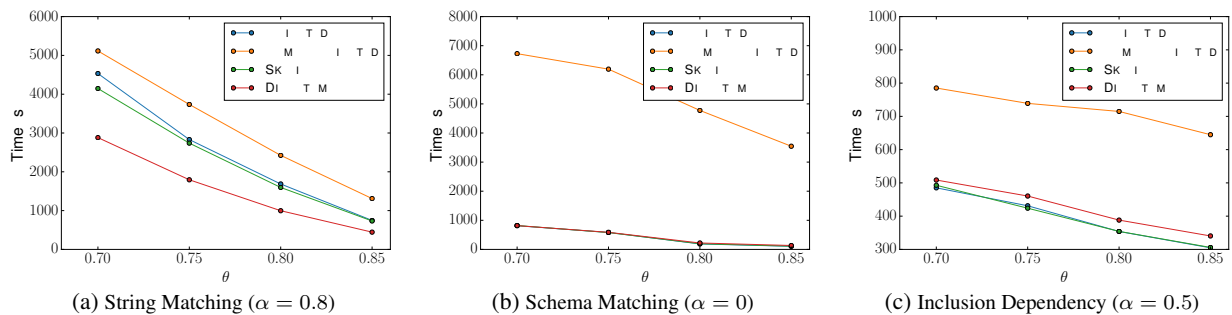

(a) String Matching ($\alpha = 0.8$)  (b) Schema Matching ($\alpha = 0$)  (c) Inclusion Dependency ($\alpha = 0.5$)

**Figure 5: Runtime performance of the signature schemes with varying $\theta$.**

each signature scheme, we generate the "best" signature for a set based on the greedy algorithms introduced in the earlier sections.

Figure 5 shows the runtime performance of the signature schemes with varying $\theta$. Note for RELATED SET SEARCH, the reported runtimes do not include the time taken to build the inverted index, while for RELATED SET DISCOVERY, the index build time was included. The number of candidates selected by these signature schemes were directly proportional to the runtimes. As expected, the signature schemes performed better with larger $\theta$; as $\theta$ increased, the threshold for relatedness became more strict and fewer candidates were considered. We saw that signatures from SKYLINE and DICHOTOMY consistently outperformed signatures from the competing COMBUNWEIGHTED for all $\theta$ and $\alpha$. Interestingly, even WEIGHTED, which did not consider the similarity threshold, also outperformed COMBUNWEIGHTED in several configurations. In the case of the schema matching application shown in Figure 5b, the signatures from WEIGHTED, SKYLINE, and DICHOTOMY had runtimes of 813, 820, and 816 seconds respectively compared to the 6276 seconds incurred by the signature from COMBUNWEIGHTED for $\theta = 0.7$, a $7.7\times$ improvement in runtime. The runtimes for WEIGHTED, SKYLINE, and DICHOTOMY were all very similar for this application because the three signature schemes all reduce to the weighted signature scheme when $\alpha = 0$. Finally, we found that DICHOTOMY performed better than SKYLINE as shown in Figure 5a when $\alpha$ was larger, and SKYLINE performed better than DICHOTOMY when $\alpha$ was smaller as shown in Figure 5c. When building the signature, if DICHOTOMY removed a token from an element, it also tended to remove the remaining tokens from the element as well due to its cost-metric; this was generally the correct thing to do when $\alpha$ was larger, but when $\alpha$ was smaller, this often led to suboptimal signatures.

### 8.3 Refinement

*Summary:* SILKMOTH's filtering mechanisms in the refinement step provided large benefits to the runtime performance, with improvements of up to $113\times$.

We evaluated the two filters introduced in Section 5 and compared them against running SILKMOTH with no filter. In these experiments, we use NOFILTER to denote no filter, CHECK to denote just the check filter, and NEARESTNEIGHBOR to denote both the check and nearest neighbor filters[13]. For all instances we ran SILKMOTH with the DICHOTOMY signature scheme and no reduction-based verification.

Figure 6 shows the results of our experiments. For all $\theta$ and $\alpha$, the runtime performance of CHECK and NEARESTNEIGHBOR vastly outstripped the performance of NOFILTER. For the inclusion dependency application shown in Figure 6c, NOFILTER, CHECK, and NEARESTNEIGHBOR incurred runtimes of 509, 67, and 4.5 seconds respectively for $\theta = 0.7$, an improvement of two orders of

---

magnitude. Since NEARESTNEIGHBOR is the more powerful filter, NEARESTNEIGHBOR reduced a greater number of candidates and was unsurprisingly faster in the vast majority of experiments.

### 8.4 Reduction-Based Verification

*Summary:* Overall, the reduction optimization offered a $30 - 50\%$ improvement in runtime.

We ran SILKMOTH with and without the reduction optimization, introduced in Section 5.3 for the inclusion dependency application. The reduction optimization was particularly suited for this application since the sets in this application had more elements on average, and the maximum matching step was particularly expensive. For this experiment, only sets with at least 100 elements were considered for the randomly chosen 1000 reference sets. We fixed $\alpha = 0$, since the reduction optimization is only valid when $\alpha = 0$. We ran SILKMOTH with the DICHOTOMY signature scheme and the NEARESTNEIGHBOR filter. The results are presented in Figure 7. Reduction gave large advantages in runtime for every $\theta$. For $\theta = 0.75$, SILKMOTH with no reduction ran in 105 seconds, whereas with reduction, it ran in 58 seconds, marking a 45% improvement in runtime. For string matching and schema matching, the advantage of reduction was minimal since the average size of the sets in these problems were very small (9 and 3 respectively).

### 8.5 Comparison Against Existing Work

*Summary:* SILKMOTH is faster than FASTJOIN at the string matching application by up to $13\times$.

We compared SILKMOTH with all its optimizations against the existing FASTJOIN work developed by Wang et al. [25]. FASTJOIN was developed as a solution to the approximate string matching algorithm. It does not target the general set relatedness, could not support the SET-CONTAINMENT metric, and only supports edit similarity for its similarity function. Therefore, we only compared SILKMOTH against FASTJOIN for the string matching application.

Figures 8a and 8b present the runtimes of both systems for varying $\theta$ and $\alpha$. SILKMOTH outperformed FASTJOIN in almost every experimental configuration. In particular, for $\theta = 0.8$ and $\alpha = 0.7$, FASTJOIN incurred a runtime of 10320 seconds, while SILKMOTH only incurred a runtime of 796 seconds, a runtime improvement of $13\times$. Only when $\alpha$ was very large, did we see that FASTJOIN was comparable to SILKMOTH. The primary reason is due to FASTJOIN's use of partitions as tokens, as opposed to SILKMOTH's use of $q$-grams. It has been established that partitions typically tend to perform better than $q$-grams for string matching applications [18]. However, we believe that many of our techniques, such as the filtering mechanisms in the refinement step, can also be applied to using partitions as tokens.

### 8.6 Scalability

Finally, we evaluated SILKMOTH's ability to scale with the input data size. We varied the number of sets in each application and ran SILKMOTH with all its optimizations. Figure 9 gives the results.

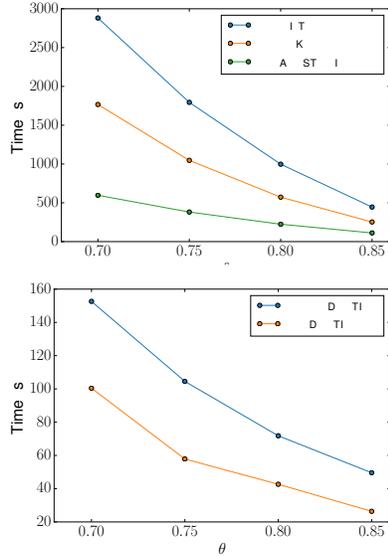
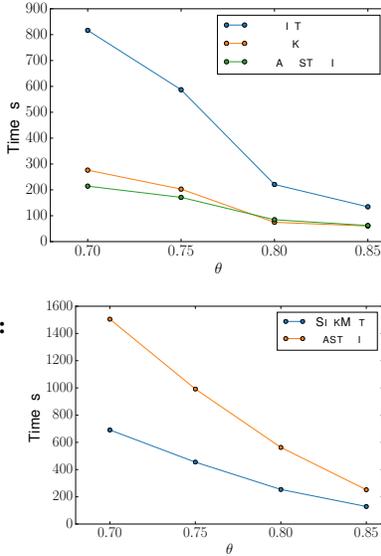
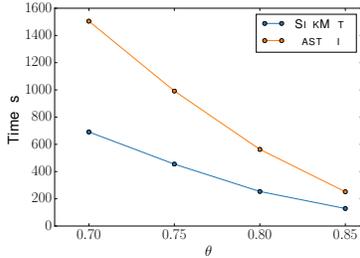
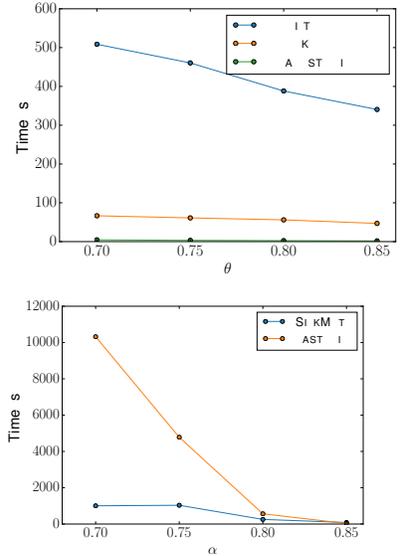

**Figure 7: Inclusion Dependency ($\alpha = 0$)**    **Figure 8: String matching for varying $\theta$ ($\alpha = 0.8$) and varying $\alpha$ ($\theta = 0.8$).**

Overall, SILKMOTH scales well. For example, the schema matching application with the RELATED SET DISCOVERY problem took 68 and 1993 seconds for 500K and 2.5M sets respectively.

## 9. RELATED WORK

There are many proposals for finding similar sets from a collection of sets [2,16,22,28]. However, most of them cannot tolerate any sort of approximate matching between elements. The most relevant technique to our work in this field is the prefix filter [4,28], which SILKMOTH adapts to efficiently find the nearest neighbors. Related work in this field includes that of Bayardo et al. [4], who proposed using prefix filters to solve the set similarity join problem. Xiao et al. [28,29] improved the prefix filtering by considering the token positions, and Deng et al. [8] extended the prefix filter for the string similarity join.

Several papers have focused tolerating a small error between the elements in set matching problems. Agrawal et al. [1] proposed an approximate set containment metric which takes token transformations (e.g., synonyms and abbreviations) into account and used it to retrieve sets with approximate containment relationships. They also calculated the set containment score by the maximum matching. However they still only considered identical elements or elements that match according to their pre-specified set of element transformations. Chaudhuri et al. [6] proposed a fuzzy match similarity function, which viewed a string as a sequence of tokens and evaluated the similarity as the minimum cost of 'transforming' one string to another by replacement, insertion, and deletion of tokens. Note the cost of replacing two tokens is proportional to their edit distance. SILKMOTH is different from these two works as SILKMOTH tolerates a different kind of error between elements. Wang et al. [25] proposed an approximate Jaccard similarity metric to evaluate the similarity between two strings; as noted in throughout the paper, this work is both less general than SILKMOTH, and performs up to an order of magnitude worse.

There are also many papers [5,15,24,30] propose approximate algorithms for set similarity search and join. Indyk et al. [13,15] introduced the Locality Sensitive Hashing (LSH) technique for approximate nearest neighbor search. The LSH technique uses multiple hash functions (such as MinHash [5]) to hash the sets and guarantees the similar sets are more likely to be allocated to the same bucket. Satuluri et al. [24] proposed to use LSH for candidate pruning and similarity estimation for approximate set similarity join. Zhai et al. [30] proposed to use the LSH for set similarity search with very low thresholds. All of these papers employ approximate algorithms, which are not guaranteed to find all matches, in contrast to SILKMOTH, which uses exact algorithms and finds all matches.

## 10. CONCLUSION

We presented SILKMOTH, a general-purpose related set discovery system. We formalized the related set discovery and search problems under various relatedness and similarity metrics. Furthermore, we extensively analyzed the optimal signature generation problem, showed it was NP-Complete, and characterized the space of valid signatures. We proposed several novel optimizations which greatly increased the performance of SILKMOTH in practice on real datasets, allowing it to outperform the existing method FASTJOIN by an order of magnitude while providing a more general set of capabilities in terms of similarity metrics and applications.

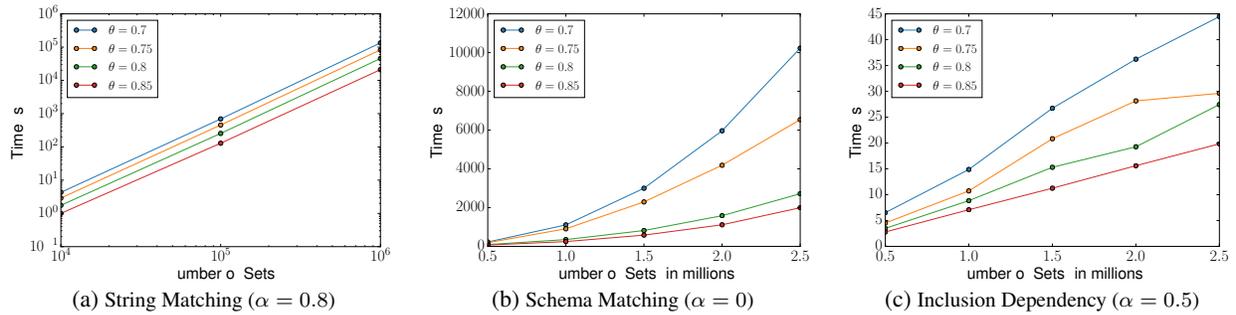

(a) String Matching ($\alpha = 0.8$)  (b) Schema Matching ($\alpha = 0$)  (c) Inclusion Dependency ($\alpha = 0.5$)

**Figure 9: Scalability of SILKMOTH with varying $\theta$.**

# Appendix

In this section, we formally prove the optimal weighted signature selection problem is NP-Complete. To accomplish this, we first reduce the 3-CNF-SAT problem to an inverse-prime subset sum problem, which given a number $s$ and a multi-set $\mathcal{A}$ of numbers, finds a subset of $\mathcal{A}$ whose sum is exactly $s$. Note all the numbers in $\mathcal{A}$ are in the form $\sum_{p \in P_i} 1/p$ where $P_i \subseteq P$, $P = \{p_1, p_2, \cdots, p_l\}$, $p_i$ is the $(i+3)^{th}$ prime (i.e., $p_1 = 7, p_2 = 11, \cdots$), and $l$ is an integer. More formally:

**Definition 12 (Inverse-Prime Subset Sum Problem).** *Given a number $s$ and a multi-set $\mathcal{A}$ of numbers in the form of $\sum_{p \in P'} \frac{1}{p}$ where $P' \subseteq P$, the inverse-prime subset sum problem $\langle \mathcal{A}, s, l \rangle$ returns true if and only if there exists a subset of $\mathcal{A}$ such that the sum of its elements is equal to $s$.*

Then, we reduce the inverse-prime subset sum problem to the decision version of our optimal weighted signature selection problem. The decision version of the optimal valid signature selection problem is stated as follows.

**Problem 5 (The Decision Problem).** *Given a set $R$, an inverted index $\mathcal{I}$, a relatedness threshold $\delta$, and an integer $k$, the decision problem $\langle \mathcal{I}, R, \delta, k \rangle$ returns true if and only if there exists a valid signature $K_R^{\mathcal{T}}$ such that $\sum_{t \in K_R^{\mathcal{T}}} |\mathcal{I}[t]| \leq k$.*

The inverse-prime subset sum problem is NP-Complete (NPC) as formalized in Lemma 3. We reduce the 3-CNF-SAT problem to this problem. A 3-CNF formula $\phi$ consists of boolean variables $x_1, x_2 \cdots, x_n$ and clauses $c_1, c_2, \cdots, c_m$. Each clause contains exactly three literals. A literal is either a variable itself $x_i$ or its negation $\overline{x_i}$. The 3-CNF-SAT problem asks whether there exists an assignment to the variables such that the formula $\phi$ is satisfied.

**Lemma 3.** *The inverse-prime subset sum problem is NP-Complete.*

**Proof.** First, given a subset of $\mathcal{A}$, we can easily check whether their sum is equal to $s$ in polynomial time.

Next we prove the inverse-prime subset sum problem is NPC by a reduction from the 3-CNF-SAT problem. In particular, given any instance $\phi$ of a 3-CNF-SAT problem, we construct an instance of the inverse-prime subset sum problem $\langle \mathcal{A}, s, l \rangle$ such that the 3-CNF formula $\phi$ is satisfied if and only if there exists a subset of $\mathcal{A}$ whose sum is equal to $s$. In the reduction, we let $l = n + m$ and $s = \sum_{i=1}^{n} 1/p_i + 3 \sum_{i=n+1}^{n+m} 1/p_i$ where $n$ and $m$ are respectively the numbers of variables and clauses in $\phi$. For example, consider the 3-CNF formula:

$$\phi = (x_1 \vee x_2 \vee x_3) \wedge (\overline{x_1} \vee \overline{x_2} \vee x_3) \wedge (\overline{x_1} \vee x_2 \vee \overline{x_3}) \wedge (x_1 \vee \overline{x_2} \vee x_3)$$

We have $n = 3$ and $m = 4$. Thus we set $l = n + m = 7$ and

$$s = \frac{1}{7} + \frac{1}{11} + \frac{1}{13} + 3 \left( \frac{1}{17} + \frac{1}{19} + \frac{1}{23} + \frac{1}{29} \right)$$

The construct of $\mathcal{A}$ is as follows. We first construct a set of "true" primes $P_t$ and a set of "false" primes $P_f$ for each variable $x_i$ in $\phi$ in the following way:

- Add $p_i$ to both $P_t$ and $P_f$
- Add $p_{j+n}$ to $P_t$ if $x_i$ is in clause $c_j$ for $1 \leq j \leq m$
- Add $p_{j+n}$ to $P_f$ if $\overline{x_i}$ is in clause $c_j$ for $1 \leq j \leq m$

Then we add the two numbers $t_i = \sum_{p \in P_t} 1/p$ and $f_i = \sum_{p \in P_f} 1/p$ to $\mathcal{A}$.

Following the previous example, for $x_1$, as $x_1$ is in clauses $c_1$ and $c_4$, we have $P_t = \{p_1, p_4, p_7\}$ and

$$t_1 = \frac{1}{p_1} + \frac{1}{p_4} + \frac{1}{p_7} = \frac{1}{7} + \frac{1}{17} + \frac{1}{29}.$$

In addition, as $\overline{x_1}$ is in clauses $c_2$ and $c_3$, we have $P_f = \{p_1, p_5, p_6\}$ and

$$f_1 = \frac{1}{p_1} + \frac{1}{p_5} + \frac{1}{p_6} = \frac{1}{7} + \frac{1}{19} + \frac{1}{23}.$$

Similarly, we can add $t_2$, $f_2$, $t_3$ and $f_3$ as shown in Table 4 to $\mathcal{A}$.

**Table 4: The constructed numbers $t_i$ and $f_i$ in $\mathcal{A}$.**

| | | i=1 | i=2 | i=3 | j=1 | j=2 | j=3 | j=4 |
|---|---|---|---|---|---|---|---|---|
| Number | | 1/7 | 1/11 | 1/13 | 1/17 | 1/19 | 1/23 | 1/29 |
| $t_1$ | | 1 | 0 | 0 | 1 | 0 | 0 | 1 |
| $f_1$ | | 1 | 0 | 0 | 0 | 1 | 1 | 0 |
| $t_2$ | | 0 | 1 | 0 | 1 | 0 | 1 | 0 |
| $f_2$ | | 0 | 1 | 0 | 0 | 1 | 0 | 1 |
| $t_3$ | | 0 | 0 | 1 | 1 | 1 | 0 | 1 |
| $f_3$ | | 0 | 0 | 1 | 0 | 0 | 1 | 0 |

Then, for each clause $c_j$ in $\phi$, we add two numbers $u_j = 1/p_{n+j}$ and $v_j = 1/p_{n+j}$ to $\mathcal{A}$. Following previous example, we have $u_1 = v_1 = 1/p_4 = 1/17$, $u_2 = v_2 = 1/p_5 = 1/19$, $u_3 = v_3 = 1/p_6 = 1/23$, $u_4 = v_4 = 1/p_7 = 1/29$ as shown in Table 5. This finishes the construction of $\mathcal{A}$.

**Table 5: The constructed numbers $u_i$ and $v_i$ in $\mathcal{A}$.**

| | | i=1 | i=2 | i=3 | j=1 | j=2 | j=3 | j=4 |
|---|---|---|---|---|---|---|---|---|
| Number | | 1/7 | 1/11 | 1/13 | 1/17 | 1/19 | 1/23 | 1/29 |
| $u_1$ | | 0 | 0 | 0 | 1 | 0 | 0 | 0 |
| $v_1$ | | 0 | 0 | 0 | 1 | 0 | 0 | 0 |
| $u_2$ | | 0 | 0 | 0 | 0 | 1 | 0 | 0 |
| $v_2$ | | 0 | 0 | 0 | 0 | 1 | 0 | 0 |
| $u_3$ | | 0 | 0 | 0 | 0 | 0 | 1 | 0 |
| $v_3$ | | 0 | 0 | 0 | 0 | 0 | 1 | 0 |
| $u_4$ | | 0 | 0 | 0 | 0 | 0 | 0 | 1 |
| $v_4$ | | 0 | 0 | 0 | 0 | 0 | 0 | 1 |

Obviously, the time complexity of this reduction is $\mathcal{O}((n+m)^2)$ as it constructs $2(n+m)$ numbers and each number consists of at most $n+m$ inverse-primes, which is in polynomial time. We next show that the formula is satisfiable if and only if there exists a subset of $\mathcal{A}$ whose sum is $s$.

First suppose there exists a true assignment that satisfies $\phi$. We choose a subset of $\mathcal{A}$ as follows.

- If $x_i$ is true, pick $t_i$
- If $x_i$ is false, pick $f_i$
- If the number of true literals in $c_j$ is at most 2, pick $u_j$
- If the number of true literals in $c_j$ is 1, pick $v_j$.

Since $x_i$ is either true or false, one and only one of $t_i$ and $f_i$ is picked for all $1 \leq i \leq n$. Thus, there is exactly 1 term $1/p_i$ in the sum of the subset. Moreover, to satisfy the formula $\phi$, for each clause $c_j$, at least one literal in $c_j$ is true. On the other hand, there are no more than 3 literals in each clause. Based on the selection of the subset, there are exactly 3 terms $1/p_{j+n}$ in the sum of the subset for all $1 \leq j \leq m$. Thus, the sum of the selected subset is exactly $s$. Following previous example, as the assignment of $x_1 = x_2 = x_3 = true$ satisfies $\phi$, we have the subset of $\mathcal{A}$ as $\{t_1, t_2, t_3, u_2, u_3, u_3, v_3, u_4\}$ as shown in Table 6.

**Table 6: The subset of $\mathcal{A}$ whose sum is $s$.**

| | | i=1 | i=2 | i=3 | j=1 | j=2 | j=3 | j=4 |
|---|---|---|---|---|---|---|---|---|
| Number | | 1/7 | 1/11 | 1/13 | 1/17 | 1/19 | 1/23 | 1/29 |
| $t_1$ | | 1 | 0 | 0 | 1 | 0 | 0 | 1 |
| $t_2$ | | 0 | 1 | 0 | 1 | 0 | 1 | 0 |
| $t_3$ | | 0 | 0 | 1 | 1 | 1 | 0 | 1 |
| $u_2$ | | 0 | 0 | 0 | 0 | 1 | 0 | 0 |
| $v_2$ | | 0 | 0 | 0 | 0 | 1 | 0 | 0 |
| $u_3$ | | 0 | 0 | 0 | 0 | 0 | 1 | 0 |
| $v_3$ | | 0 | 0 | 0 | 0 | 0 | 1 | 0 |
| $u_4$ | | 0 | 0 | 0 | 0 | 0 | 0 | 1 |
| $s$ | | 1 | 1 | 1 | 3 | 3 | 3 | 3 |

Next suppose that there exists a subset of $\mathcal{A}$ whose sum is $s$. We prove there must exists an assignment that satisfies $\phi$. We add up all the numbers in the subset. Suppose we only combine the fractions in the numbers with the same denominator and have $\sum_{i=1}^{n+m} b_i/p_i$ where $b_i$ is an non-negative integer. Next we prove that $b_i$ can only be 1 for all $1 \leq i \leq n$ and $b_j$ can only be 3 for all $n+1 \leq j \leq n+m$.

Since the sum of the subset is $s$, we have

$$\sum_{i=1}^{n+m} b_i/p_i = \sum_{i=1}^{n} 1/p_i + 3 \sum_{i=n+1}^{n+m} 1/p_i \qquad (6)$$

$$\prod_{j=1}^{n+m} p_j \sum_{i=1}^{n+m} b_i/p_i = \prod_{j=1}^{n+m} p_j \left( \sum_{i=1}^{n} 1/p_i + 3 \sum_{i=n+1}^{n+m} 1/p_i \right) \quad (7)$$

$$\sum_{i=1}^{n+m} b_i \prod_{j \neq i}^{n+m} p_j = \sum_{i=1}^{n} \prod_{j \neq i}^{n+m} p_j + 3 \sum_{i=n+1}^{n+m} \prod_{j \neq i}^{n+m} p_j. \quad (8)$$

Next for each $1 \leq k \leq n$, we modulo $p_k$ on both sides of the above equation. Since all the terms with $p_k$ would get a remainder of 0 and we have

$$b_k \prod_{j \neq k}^{n+m} p_j \mod p_k = \prod_{j \neq k}^{n+m} p_j \mod p_k \qquad (1)$$

$$(b_k - 1) \prod_{j \neq k}^{n+m} p_j \mod p_k = 0. \qquad (2)$$

Based on the construction of $\mathcal{A}$, $1 \leq b_k \leq 2$. $p_i$ must be a prime for all $1 \leq i \leq n+m$, so $b_k$ must be 1 to satisfy the above equation. Similarly, for any $n+1 \leq k \leq n+m$ we have

$$(b_k - 3) \prod_{j \neq k}^{n+m} p_j \mod p_k = 0.$$

Based on the construct of $\mathcal{A}$, $1 \leq b_k \leq 5$. Furthermore, $p_i$ must be prime and larger than 5 for all $1 \leq i \leq n+m$, so $b_k$ can only be 3 to satisfy the above equation.

Then we assign the value true to $x_i$ if $t_i$ is in the subset. We assign the value false to $x_i$ if $f_i$ is in the subset. There is exactly one value per variable in the subset since $b_i = 1$ for all $1 \leq i \leq n$. Moreover, since $b_j = 3$ for all $n+1 \leq j \leq n+m$, there is at least one true literal in each clause. Thus this assignment satisfies $\phi$ and the theorem is proved. $\square$

**Theorem 6.** *The decision problem is NP-Complete.*

Proof. First, give a signature $K_R^{\mathcal{T}}$, we can easily check whether it is a valid signature in weighted signature scheme and whether $\sum_{t \in K_R^{\mathcal{T}}} |\mathcal{I}[t]| \leq k$ in polynomial time.

Next we prove the decision problem is NPC by a reduction from the inverse-prime subset sum problem. In particular, given any instance $\langle \mathcal{A}, s, l \rangle$ of an inverse-prime subset sum problem, we construct a decision problem $\langle \mathcal{I}, R, \delta, k \rangle$ such that there exists a subset of $\mathcal{A}$ whose sum is $s$ if and only if there is a valid signature $K_R^{\mathcal{T}}$ such that $\sum_{t \in K_R^{\mathcal{T}}} |\mathcal{I}[t]| \leq k$.

Suppose the numbers in $\mathcal{A}$ are respectively $a_1, a_2, \cdots, a_{|\mathcal{A}|}$ where

$$a_i = \sum_{p \in P_i} \frac{1}{p},$$

$P_i \subseteq P$ and $P = \{p_1, p_2, \cdots, p_l\}$. In the reduction, for each number $a_i$, we create 1 token $t_i$ where

$$|\mathcal{I}[t_i]| = a_i \prod_{p \in P} p$$

We also create $|P_i|$ elements, one $r_i^p$ for each $p \in P_i$ where $r_i^p$ contains the token $t_i$ and another $p-1$ dummy tokens whose inverted list lengths are arbitrarily large. This finishes the construction of $\mathcal{I}$ and $R$ (which contains all the constructed elements). In addition, we let

$$k = s \prod_{p \in P} p$$

and

$$\delta = 1 - \frac{s - \varepsilon}{\sum_{a_i \in \mathcal{A}} |P_i|}$$

where $\varepsilon$ is an arbitrarily small positive number. Note when $s = 0$, the inverse-prime subset sum problem is trivial and always finds the empty set whose sum is $s = 0$. Similarly when

$$s > \sum_{a_i \in \mathcal{A}} |P_i| = \sum_{a_i \in \mathcal{A}} \sum_{p \in P_i} 1 \geq \sum_{a_i \in \mathcal{A}} \sum_{p \in P_i} 1/p = \sum_{a_i \in \mathcal{A}} a_i,$$

the inverse-prime subset sum problem is trivial and always cannot find a subset of $\mathcal{A}$ whose sum is $s$, as the sum of all the numbers in $\mathcal{A}$ is smaller than $s$. For the other cases, we have $\delta \in [0, 1]$.

The time complexity for the reduction is

$$\mathcal{O}\left(\sum_{i=1}^{|\mathcal{A}|} \sum_{p \in P_i} p\right)$$

which is in polynomial time since $p_n = \mathcal{O}(n^2)$.[14] Note the problem size in bits is polynomial in $n$ as

$$\mathcal{O}\left(\prod_{p \in P} p\right) = \mathcal{O}\left(\prod_{i=1}^{n} p_i\right) = \mathcal{O}(n^{2n}) = \mathcal{O}(2^{2n \log n}) = \mathcal{O}(2^{poly(n)})$$

Next we show that there exists a subset of $\mathcal{A}$ whose sum is $s$ if and only if the decision problem is satisfiable.

First suppose there exists a subset $\mathcal{A}'$ of $\mathcal{A}$ whose sum is $s$. We have

$$\sum_{a_i \in \mathcal{A}'} a_i = s.$$

For each number $a_i$ in the subset $\mathcal{A}'$, we add its corresponding token $t_i$ into $K_R^{\mathcal{T}}$. Thus we have

$$\sum_{t_i \in K_R^{\mathcal{T}}} |\mathcal{I}[t_i]| = \sum_{a_i \in \mathcal{A}'} a_i \prod_{p \in P} p = s \prod_{p \in P} p = k \leq k.$$

In addition, $K_R^{\mathcal{T}}$ is a valid signature as

$$\sum_{r_i^p \in R} \frac{|r_i^p| - |r_i^p \cap K_R^{\mathcal{T}}|}{|r_i^p|} = \sum_{a_i \in \mathcal{A}'} \sum_{p \in P_i} \frac{|r_i^p| - 1}{|r_i^p|} + \sum_{a_i \in \mathcal{A} \setminus \mathcal{A}'} \sum_{p \in P_i} \frac{|r_i^p| - 0}{|r_i^p|} \quad (1)$$

$$= \sum_{a_i \in \mathcal{A}} |P_i| - \sum_{a_i \in \mathcal{A}'} \sum_{p \in P_i} \frac{1}{|r_i^p|} \quad (2)$$

$$= \sum_{a_i \in \mathcal{A}} |P_i| - \sum_{a_i \in \mathcal{A}'} \sum_{p \in P_i} \frac{1}{p} \quad (3)$$

$$= \sum_{a_i \in \mathcal{A}} |P_i| - \sum_{a_i \in \mathcal{A}'} a_i \quad (4)$$

$$= \sum_{a_i \in \mathcal{A}} |P_i| - s \quad (5)$$

$$= (1 - \frac{s}{\sum_{a_i \in \mathcal{A}} |P_i|}) \sum_{a_i \in \mathcal{A}} |P_i| \quad (6)$$

$$= (1 - \frac{s}{\sum_{a_i \in \mathcal{A}} |P_i|}) |R| \quad (7)$$

$$< (1 - \frac{s - \varepsilon}{\sum_{a_i \in \mathcal{A}} |P_i|}) |R| \quad (8)$$

$$= \delta |R|. \quad (9)$$

Note from (6) to (7), based on the construction of $R$, for each number $a_i$ in $\mathcal{A}$, we create $|P_i|$ elements for $R$. Thus

$$|R| = \sum_{a_i \in \mathcal{A}} |P_i|.$$

Next suppose there exists a valid signature $K_R^{\mathcal{T}}$ such that

$$\sum_{t_i \in K_R^{\mathcal{T}}} |\mathcal{I}[t_i]| \leq k.$$

Then for each $t_i \in K_R^{\mathcal{T}}$, we add its corresponding number $a_i$ in $\mathcal{A}$ to the subset $\mathcal{A}'$. We show $\sum_{a_i \in \mathcal{A}'} a_i = s$. On the one hand, we have

$$\sum_{t_i \in K_R^{\mathcal{T}}} |\mathcal{I}[t_i]| = \sum_{a_i \in \mathcal{A}'} a_i \prod_{p \in P} p \leq k = s \prod_{p \in P} p.$$

Thus

$$\sum_{a_i \in \mathcal{A}'} a_i \leq s.$$

On the other hand, we have

$$\sum_{r_i^p \in R} \frac{|r_i^p| - |r_i^p \cap K_R^{\mathcal{T}}|}{|r_i^p|} = \sum_{a_i \in \mathcal{A}} |P_i| - \sum_{a_i \in \mathcal{A}'} a_i < \delta |R| = \sum_{a_i \in \mathcal{A}} |P_i| - (s - \varepsilon).$$

Thus

$$s - \varepsilon < \sum_{a_i \in \mathcal{A}'} a_i \leq s.$$

As $\varepsilon$ is an arbitrarily small positive number, we have

$$\sum_{a_i \in \mathcal{A}'} a_i = s.$$

$\square$